# Necklace-structured high harmonic generation for low-divergence, soft X-ray harmonic combs with tunable line spacing

Laura Rego[1]†, Nathan J. Brooks[2]†, Quynh L. D. Nguyen[2]*, Julio San Román[1], Iona Binnie[2], Luis Plaja[1], Henry C. Kapteyn[2], Margaret M. Murnane[2], Carlos Hernández-García[1]

[1]Grupo de Investigación en Aplicaciones del Láser y Fotónica, Departamento de Física Aplicada, University of Salamanca, Salamanca E-37008, Spain

[2]JILA - Department of Physics, University of Colorado and NIST, Boulder, Colorado 80309, USA

†These authors contributed equally to this work.

*Corresponding Authors: Quynh.L.Nguyen@colorado.edu

## Abstract

The extreme nonlinear optical process of high-harmonic generation (HHG) makes it possible to map the properties of a laser beam onto a radiating electron wavefunction, and in turn, onto the emitted x-ray light. Bright HHG beams typically emerge from a longitudinal phased distribution of atomic-scale quantum antennae. Here, we form a transverse necklace-shaped phased array of HHG emitters, where orbital angular momentum conservation allows us to tune the line spacing and divergence properties of extreme-ultraviolet and soft X-ray high harmonic combs. The on-axis HHG emission has extremely low divergence, well below that obtained when using Gaussian driving beams, which further decreases with harmonic order. This work provides a new degree of freedom for the design of harmonic combs – particularly in the soft X-ray regime, where very limited options are available. Such harmonic beams can enable more sensitive probes of the fastest correlated charge and spin dynamics in molecules, nanoparticles and materials.

# Introduction

A new generation of coherent x-ray sources are opening up a new understanding of the fastest coupled charge, spin and phonon interactions and transport in materials (1-5). X-ray sources such as free electron lasers (6, 7) and high-harmonic generation (HHG) (8-14) can produce coherent light from the extreme-ultraviolet (EUV) to the soft X-ray (SXR) region. Moreover, in the case of HHG sources, they are perfectly synchronized to the driving laser, to sub-femtosecond precision, and present high temporal coherence. In HHG, an atom undergoes strong field ionization in an intense femtosecond laser field. The liberated electron is then accelerated in the laser field before recombining with the parent ion, which results in the emission of higher-order harmonics. In the microscopic quantum picture, the driving laser creates a nanoscale dipole antenna in each atom, which radiates high harmonics of the fundamental laser field. This short wavelength radiation can be manipulated at the single atom level or at the macroscopic phase-matching level by structuring the intensity, frequency content, polarization and orbital angular momentum (OAM) of the driving laser field (15-21).

Currently, control of the frequency content of the HHG light source – a key property for many advanced applications – is gained by changing the wavelength of the driving laser or by using frequency-selective optics or monochromators. By tuning the driving laser wavelength from the mid- and near-infrared to the ultraviolet, the HHG spectrum can be tuned from a bright coherent supercontinuum to a set of narrow-band (22, 23). Bright high-order harmonics extending into the keV region, well beyond the water window, can be obtained by using mid-infrared driving fields (22, 24-30). Narrow spectral peaks into the SXR region can be achieved by driving harmonics with UV lasers, although the low ponderomotive energy ($\sim\lambda^2$) necessitates the use of extremely high laser intensities (23). The manipulation of the frequency and the divergence of these HHG beams is still challenging and demands very efficient monochromators and good focusing optics in the EUV and SXR regions. An appealing alternative is to instead imprint the desired properties directly onto the HHG light, by tailoring the driving laser and taking advantage of selection rules. Recent works have demonstrated the relevance of non-perturbative dynamics underlying HHG in rephasing the wavefront to achieve good divergence profiles (31, 32), which can lead to focusing of harmonic beams without additional optics. However, it is still not possible to achieve general/full control over the frequency content and divergence properties of HHG radiation, to tailor the HHG illumination for applications.

The emerging field of ultrafast structured light is providing exciting techniques for enhancing laser-matter interactions for applications (33). In particular, exploitation of the OAM is opening up unexpected avenues for controlling the properties of high-harmonic fields as they are being generated. OAM manifests itself as a variation of the beam's spatial phase along its transverse profile, and it is characterized by its topological charge, $\ell$, or number of $2\pi$ phase twists along the azimuthal coordinate (34, 35). Since the first experiments in 2012 (36), OAM-driven HHG has proven to be a powerful tool for shaping the spatial properties of higher order harmonics—including the topological charge, intensity distribution (37-42), and polarization properties (20, 43). OAM can also be used to control the temporal shape of HHG—through the generation of helical attosecond pulses (37,39) or high harmonics with an OAM that increases during the pulse, which is a unique capability of HHG light (21). Although the ability to control the temporal shape implies an ability to also control the spectral shape, to date OAM has not been yet exploited to tailor the spectral content in HHG.

In this work, we are able to control the spectral and divergence properties of HHG by driving it with optimally phased necklace laser beams – a class of ring-shaped beams with azimuthally modulated amplitude and phase, resulting from the interference of multiple OAM modes. Our theoretical and experimental results show that by driving HHG with these structured beams, we generate a transverse phased array of HHG sources which emit a bright and adjustable harmonic comb along the optical axis. OAM selection rules and transverse phase-matching conditions allow us to tune the spectral spacing and divergence properties of these harmonics. Significantly, the on-axis HHG emission has extremely low divergence, which is not only lower than that obtained with standard Gaussian beams, but the scaling behavior with the harmonic order is also reversed — such that higher-order harmonics exhibit progressively lower divergences (see Fig. 1). Our simulations demonstrate that these properties extend into the water window SXR regime, when driven by mid-IR necklace beams. This work provides a new degree of freedom for the design of harmonic combs – particularly in the SXR regime, where very limited options are available.

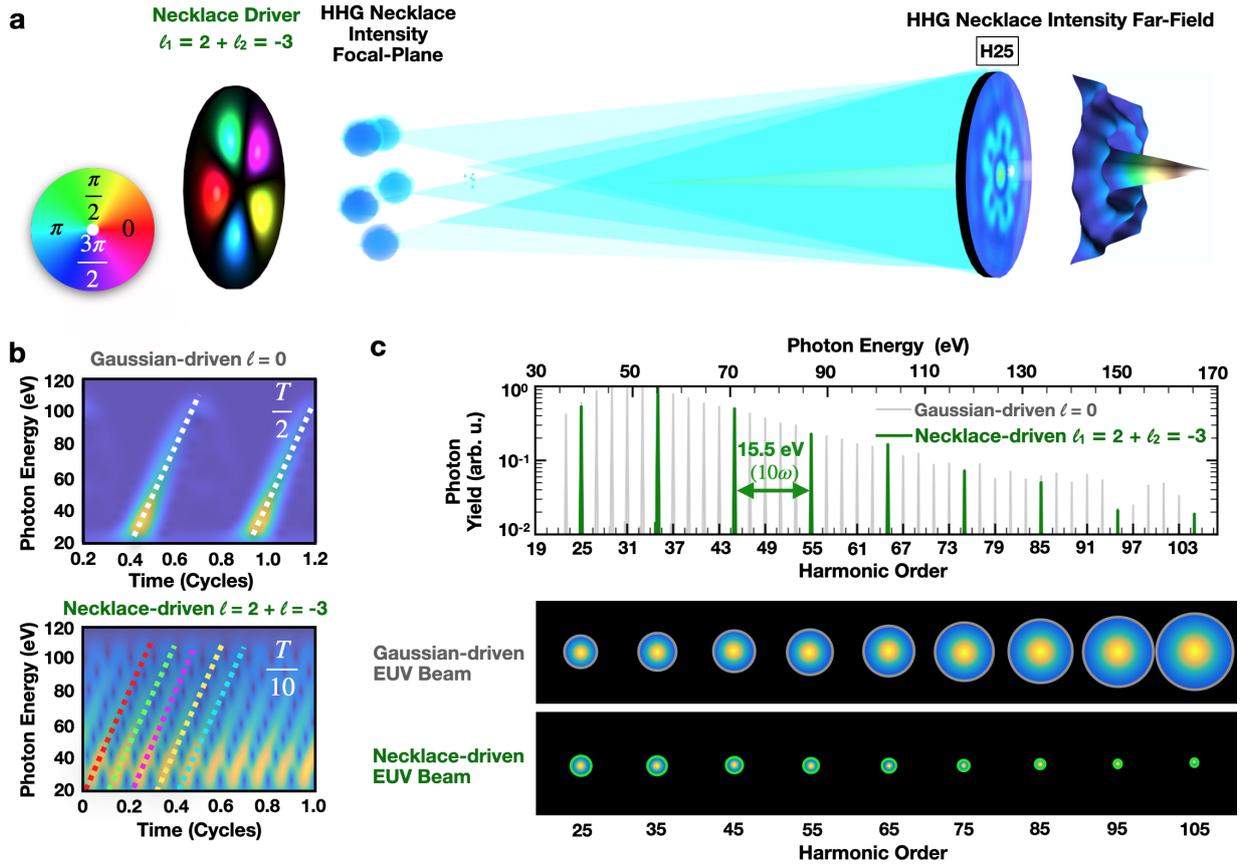

**Figure 1 | Tunable, low-divergence high harmonic combs via necklace-driven HHG. a** Two linearly polarized vortex beams carrying OAM of $\ell_1=2$ and $\ell_2=-3$ are overlapped to create a necklace-structured intensity and phase profile, and focused into He/Ar gas to drive HHG. The intensity lobes at the necklace focal plane represent a phased array of EUV/SXR emitters which interfere on-axis to form a comb of harmonics with a spacing dependent on the OAM of the driving fields. The harmonic intensity profile shows a strong emission on-axis (detailed in the figure for the 25$^{th}$ harmonic). **b** Time-frequency analysis of the simulated on-axis harmonic comb for Gaussian and necklace ($\ell_1=2$, $\ell_2=-3$) driven HHG in He at 800 nm. The dashed-color lines indicate the corresponding phase from panel **a** for each emission event. **c** HHG spectrum for the case of necklace-driven and Gaussian-driven HHG simulated in panel **b**. The harmonic spacing, $\Delta\omega=10\omega_0$, (15.5 eV), in the necklace-driven on-axis HHG spectrum is a result of OAM conservation, and is tunable by varying the OAM content of the driving laser (upper panel). The on-axis harmonics are emitted with a divergence which is significantly reduced compared to that of Gaussian-driven HHG, and decreases with increasing harmonic order (bottom panel).

# Tailoring the high-harmonic line spacing through necklace-driven HHG

Necklace beams can be generated and maintained through self-focusing in nonlinear media (44, 45), or by creating them utilizing either spatial light modulators (46) or specially tailored diffractive optical elements (47, 48). Here, we form necklace beams with a distinctive phase structure by spatiotemporally overlapping two femtosecond laser pulses with identical duration, wavelength, and polarization, but opposite and non-degenerate OAM ($\ell_1=|\ell_1|$, $\ell_2=-|\ell_2|$). The composite electric field exhibits a modulated intensity necklace structure, with evenly-spaced lobes of similar amplitude arranged at a constant distance from the optical axis, and with a relative phase shift between neighboring lobes. Figure 1a shows the intensity-modulated phase profile at the focus resulting from the superposition of two vortex beams with $\ell_1=2$, $\ell_2=-3$, as well as the far-field spatial intensity distribution of a given harmonic order (the $25^{th}$). Remarkably, even though the driving field contains a singularity and hence has zero intensity at all points along the optical axis, we observe numerically and experimentally that a subset of harmonic orders develops a bright on-axis maximum upon propagation. We further find that by using different combinations of OAM beams to synthesize the necklace driver, this subset can be varied, allowing us to tune the line spacing of the harmonic comb emitted on the optical axis without altering the driving laser wavelength.

This surprising result can be understood by viewing the combined dual-vortex source as an EUV/SXR phased antenna array. The composition of OAM beams creates a necklace-structured electric field containing $N = |\ell_1| +|\ell_2|$ lobes equidistant from the optical axis/origin, where the relative phase offset of the fundamental field across the $n^{th}$ lobe is constant and equal to $(2\pi n|\ell_1|)/(|\ell_1| +|\ell_2|)$ (see Supplemental Material). Treating each lobe as a radiator of the $q^{th}$ order harmonic which is coherent with the driving laser, the total field at a point $\vec{r_f}$ away from the source plane is the sum of fields propagated from all the lobes in the necklace, and can be approximated by

$$E_q(\vec{r_f}, t) \propto \sum_{n=0}^{N-1} e^{i(q\omega_0 t + qk_0 d_n + 2\pi n \frac{q|\ell_1|}{|\ell_1|+|\ell_2|})} \quad [1]$$

The second phase term $qk_0 d_n$ describes the phase accumulated by the $q^{th}$ harmonic propagating a distance $d_n = |\vec{r_f} - \vec{r_{0,n}}|$ from the $n^{th}$ lobe to the observation point. Let us now consider the harmonic emission that is emitted on-axis. For all points lying along the optical axis, these $d_n = d$ are equal. Thus, the propagation amounts to a constant phase shift (independent of *n*), which we can omit without loss of generality. The final phase term can be simplified by reordering the terms in the summation, yielding (see Supplemental Material)

$$E_q(\vec{r}_{on-axis}, t) \propto e^{iq\omega_0 t} \sum_{n=0}^{N-1} e^{i2\pi n \frac{q}{\xi_1+\xi_2}} \quad [2]$$

where we have introduced $\xi_1 = L_{lcm}/|\ell_1|$ and $\xi_2 = L_{lcm}/|\ell_2|$, $L_{lcm}$ being the least common multiple of $|\ell_1|$ and $|\ell_2|$. From the above equation, we can observe that for harmonics where the quantity $q/(\xi_1 + \xi_2)$ is an integer, the emission from all lobes arrives in phase, resulting in a maximum on axis, while for all other harmonics the contributions sum to zero. In other words, for certain harmonic orders, the necklace emitters are transversely phase-matched by the high-harmonic upconversion process to interfere constructively on the optical axis. Taking into account the additional constraint that *q* must be odd due to inversion symmetry results in an HHG comb with line spacing equal to

$$\Delta\omega = 2(\xi_1 + \xi_2)\omega_0 \quad [3].$$

The temporal counterpart of this modified line spacing manifests in the periodicity of harmonic emission recorded on the optical axis. In Fig. 1b we perform a time-frequency analysis of the on-axis harmonic signal extracted from our simulation results in He at 800 nm (see theoretical methods below). We compare the harmonic emission driven by a standard Gaussian beam against that driven by a necklace beam with OAM content $\ell_1$=2, $\ell_2$=-3. For the Gaussian driving beam, harmonics are emitted every half period of the driving field, showing a periodicity of $\Delta t = T_0/2$ (where $T_0 = 2\pi/\omega_0$ is the optical cycle associated to the central frequency of the driving field), which physically corresponds to the cadence of the ionization-rescattering mechanism leading to HHG. This structure corresponds to a harmonic frequency comb composed of odd harmonics, with spacing $\Delta\omega=2\omega_0$ (see Fig. 1c). However, for the combined $\ell_1=2$, $\ell_2=-3$ OAM driving field, harmonic events are observed on-axis 10 times—i.e. $2(\xi_1+\xi_2)$—per period of the driving frequency, reflecting the coherent addition of the 5—$\xi_1+\xi_2$—unique HHG emitters in the necklace. As a consequence, the on-axis harmonic emission shows a periodicity of $\Delta t=T_0/10$ — $\Delta t=T_0/2(\xi_1+\xi_2)$—corresponding to a harmonic frequency comb with the line spacing given by Eq. (3). We emphasize that this is accomplished without altering the wavelength of the driving laser nor the microscopic dynamics. The spectral changes arise purely as a result of the macroscopic arrangement of the phased emitters.

A deeper understanding of the modified harmonic comb spacing can be gained by invoking the selection rules resulting from OAM conservation. HHG driven by two spatiotemporally overlapped OAM pulses leads to the generation of a comb of harmonics with several, non-trivial, OAM contributions (40). In particular, neglecting the intrinsic or dipole phase contributions (40), the $q$-th order harmonic order has allowed OAM channels given by $\ell_q = n\ell_1 + (q-n)\ell_2$, where $n$ is a positive integer representing the number of photons of the $\ell_1$ driver. If we apply this conservation rule to our scheme where the two drivers have opposite, non-degenerate OAM, i.e. $\ell_1=|\ell_1|$ and $\ell_2=-|\ell_2|$, we can readily observe that high-order harmonics emitted on-axis, i.e. with $\ell_q=0$, are generated if $n|\ell_1|=(q-n)|\ell_2|$. In order to extract the allowed harmonics that fulfill this condition, and thus the content of the harmonic comb emitted on-axis, we again denote $L_{lcm}$ as the least common multiple of $|\ell_1|$ and $|\ell_2|$, which fulfills $\eta L_{lcm}=n|\ell_1|=(q-n)|\ell_2|$, $\eta$ being an integer. Retaining the definitions of $\xi_1$ and $\xi_2$, the harmonic orders emitted with $\ell_q=0$ must fulfill $q = \eta(\xi_1+\xi_2)$. Taking into account that $q$ must be odd due to the inversion symmetry, $\eta$ must be odd, and the high-order harmonics that are emitted on-axis are

$$\omega_q=(2m+1)(\xi_1+\xi_2)\omega_0, \qquad [4]$$

where $m=0,1,2…$, again leading to the line spacing rule given by Eq. (3).

The line spacing Δω, is shown in Fig. 2a. in terms of $|\ell_1|$ and $|\ell_2|$. We see that the appearance of on-axis harmonics with modified spectral spacing is a necessary consequence of a fundamental conservation law for OAM in HHG, as the OAM content of the driver determines which harmonic wavelengths have an allowed $\ell_q=0$ channel. This interpretation naturally implies that the intensity ratio between the driving beams can be chosen to optimize the $\ell_q=0$ (on axis) contribution (see Supplemental Material). Note that those harmonics that are not emitted on-axis possess non-zero OAM, and though they are present in the HHG beam, they present a singularity at the center.

To verify these predictions, we performed full quantum HHG simulations including propagation using the electromagnetic field propagator (49) (see Methods), a method that was used in several previous calculations of HHG involving OAM (20, 21, 37, 40, 43). In Figs. 2b and 2c we present the on-axis HHG spectra driven in He with 800 nm and 2 μm wavelength driving fields, respectively, and for different drivers' OAM combinations: $\ell_1$=1, $\ell_2$=-2 (blue); $\ell_1$=2, $\ell_2$=-3 (green); $\ell_1$=3, $\ell_2$=-4 (yellow); and $\ell_1$=4, $\ell_2$=-5 (red). In Fig. 2b we also show the spectra obtained with a standard Gaussian beam driver (grey). The simulation results clearly show the versatility of this technique to modify the frequency content of the harmonic combs, whose line spacing can be properly varied through the choice of the drivers' OAM, from $2\omega_0$ to $18\omega_0$ for the cases presented in Fig. 2.

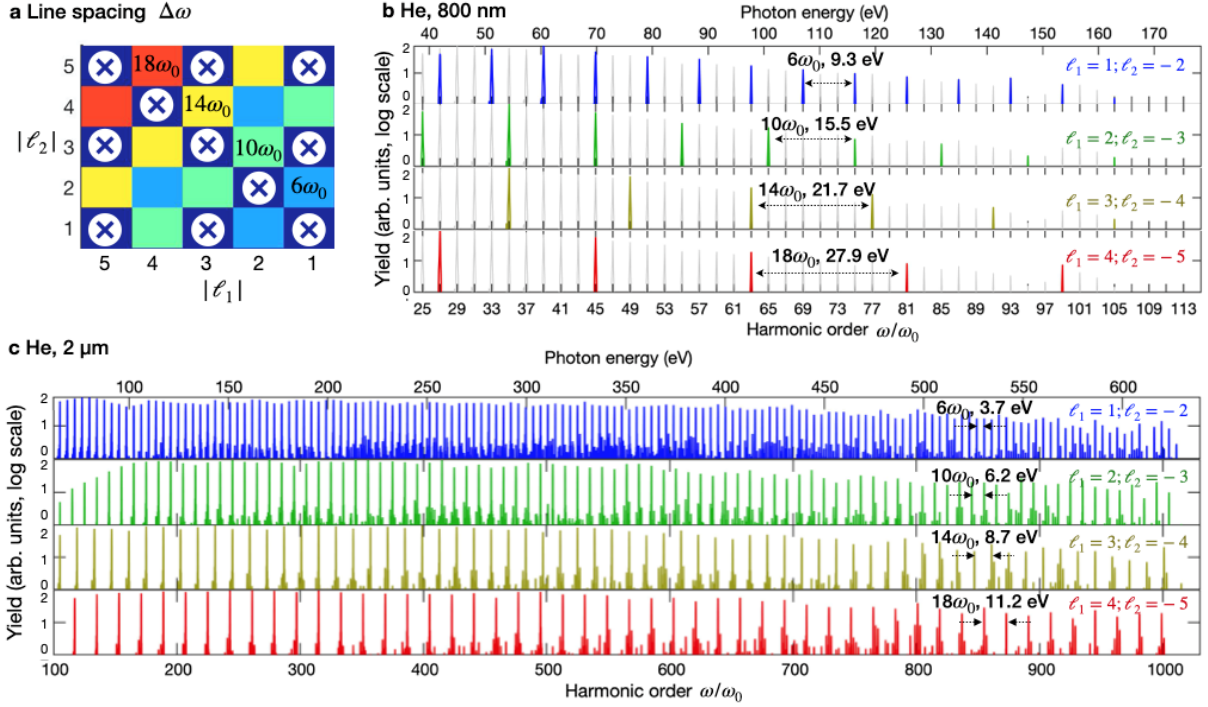

**Figure 2 | Harmonic frequency combs with tunable line spacing controllable through the drivers' OAM content.**
**a** Representation of the line spacing allowed by the selection rules for different values of $|\ell_1|$ and $|\ell_2|$. The color scale represents the line spacing, being $6\omega_0$ (blue), $10\omega_0$ (green), $14\omega_0$ (yellow) and $18\omega_0$ (red). **b, c** Simulation results of the high harmonic spectra obtained in He for **b** 800 nm and **c** 2 μm wavelength drivers respectively, for the driver's OAM combinations: $\ell_1=1$, $\ell_2=-2$ (blue); $\ell_1=2$, $\ell_2=-3$ (green); $\ell_1=3$, $\ell_2=-4$ (yellow); and $\ell_1=4$, $\ell_2=-5$ (red). The line spacing corresponds to that predicted in panel **a**. The driving beam waists of the different OAM modes are chosen to overlap at the radius ($30/\sqrt{2}$ μm) of maximum intensity ($6.9 \times 10^{14}$ W/cm² for 800 nm, and $5 \times 10^{14}$ W/cm² for 2 μm) at the focal plane. The laser pulses are modeled with a trapezoidal envelope with 26.7 fs of constant amplitude.

By using mid-infrared drivers, this technique can be exploited to customize harmonic combs extending into the SXR—as depicted in Fig. 2c, where photon energies of up to 640 eV are reached using 2 μm wavelength. This is particularly promising in a regime where, driven by long-wavelength Gaussian beams, neighboring harmonic orders tend to merge into a near or true supercontinuum (22, 23), necessitating the use of lossy dispersive optics for spectroscopic applications. In contrast, these simulation results demonstrate that, using necklace-driven HHG, the discrete and tunable peak structure can be preserved up to the SXR when driven by mid-IR wavelengths. Note that the intensity ratio between the two driving pulses has been adjusted independently for each OAM combination, in order to optimize the $\ell_q=0$ contribution (see Supplemental Material). In addition, the driving beam waists of the different OAM modes are chosen so their rings of maximum intensity overlap at $\rho_1=30/\sqrt{2}=21.21$ μm at the focal plane (see

Methods), corresponding to the ring of maximum intensity of a vortex beam with $\ell=1$ with $w_0=30$ μm, used as a reference.

To experimentally confirm the predicted spectral properties of this unique EUV light source, we use a Mach-Zehnder interferometer to synthesize the necklace-structured driver from two OAM laser beams with identical wavelength $\lambda=790$ nm and distinct topological charges of $\ell_1=1$, $\ell_2=-2$ (see Methods). The component beams are overlapped in time and space, and focused into an argon gas jet to drive HHG. The spectrum and shape of the emitted harmonics are then analyzed via a 2D imaging spectrometer consisting of a toroidal mirror and flat grating. As the on-axis HHG beam component is predicted to develop through propagation away from the source, we place an EUV charge-coupled device (CCD) camera slightly behind the focal plane of the toroidal mirror. The measured spatio-spectral images thus simultaneously record the harmonic photon energies and individual far field spatial profiles.

The measured spatial intensity profile of each of the high-order harmonics presents a structure with a symmetry similar to that of the driving necklace beam, in agreement with the results from our theoretical simulations (see Fig. 3). In this OAM combination, the predicted line spacing of the on-axis harmonic comb is $\Delta\omega=6\omega_0$, (9.5 eV) and the highest measurable harmonic orders allowed by the selection rule given by Eq. (2) are the 15th (23.5 eV) and the 21st (33.0 eV). This is supported by the intensity profiles shown in Fig. 3, where the central bright spot in the 15th and the 21st harmonics indicates a strong on-axis ($\ell=0$) component. In contrast, all other observed harmonic orders exhibit a central null, i.e., they only possess non-zero OAM contributions which lead to off-axis emission profiles. In order to confirm that these features indeed constitute the predicted harmonic combs with on-demand line spacing, we insert a small circular aperture into the HHG beam prior to the spectrometer. We observe that the harmonics with on-axis components are cleanly transmitted through the aperture, while other harmonic orders are strongly suppressed. In the Supplemental Material, experimental results for another phased-necklace driver ($\ell_1=2$, $\ell_2=-$

3) further support the predicted line spacing of the on-axis harmonic combs.

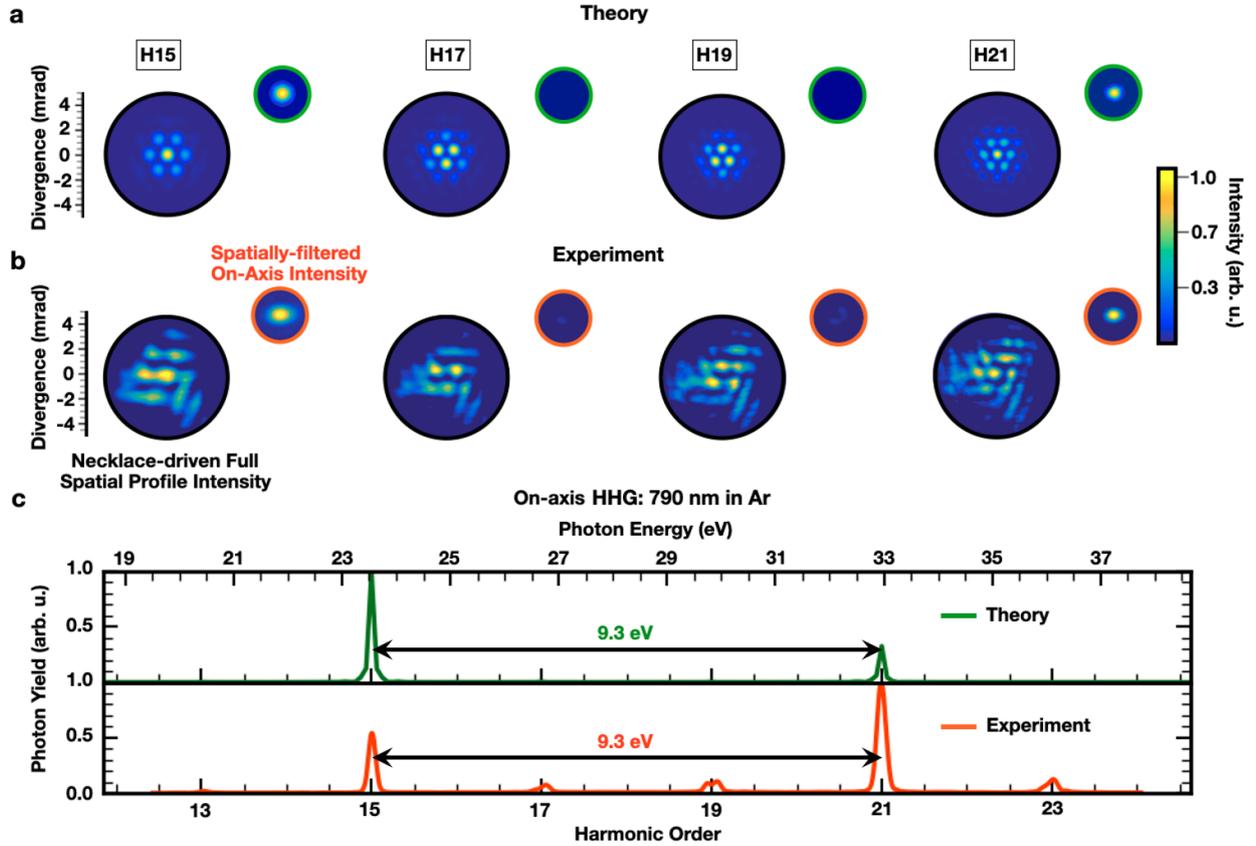

**Figure 3 | Experimental and theoretical high harmonic combs in Ar gas using a pair of 790-nm OAM driving lasers with opposite parity** ($\ell_1 = 1$ and $\ell_2 = -2$). The intensity spatial structure for the 15$^{th}$ (H15) to 21$^{st}$ (H21) harmonics are shown for **a** theory and **b** experiment. On-axis emission is allowed for H15 and H21 as a result of OAM selection rules and transverse phase matching conditions. Thus, the on-axis component is transmitted for H15 and H21, while H17 and H19 are blocked (insets in **a** & **b**). **c** Simulated (top) and experimental (bottom) HHG spectra of necklace-driven on-axis emission. By measuring the spectrum along a line in the dispersion plane which intersects the optical axis for all orders, we find that the line spacing of the transmitted harmonics, $\Delta\omega=6\omega_0$, (9.5 eV), is consistent with that predicted by OAM conservation laws. Small HHG signals experimentally observed at other harmonic orders are due to leak-through of the components carrying higher topological charges, and could be further suppressed by using a smaller aperture. The difference in the ratio H15/H21 is due to slightly different cutoff energies between simulation and experiment.

## **Generation of low-divergence high-order harmonics via necklace-driven HHG**

In addition to their spectral content, the divergence of the high harmonic combs is also crucial for applications in x-ray spectroscopy and imaging. In our scheme, the on-axis harmonics with on-demand line spacing are generated with a remarkably low divergence, compared to that obtained from standard Gaussian driving laser beams. This is a consequence of the generation mechanism, in which the on-axis beam, not present in the source plane, arises from the constructive interference of multiple phased-shifted radiators arranged with circular symmetry about the optical axis. Figure 4a shows the simulation results of the spatial intensity profile (top) and OAM content (bottom) of the 27$^{th}$ harmonic generated in He at 800 nm for two different OAM combinations: $\ell_1$=1, $\ell_2$=-2 (left) and $\ell_1$=4, $\ell_2$=-5 (right). It can be observed that most of the harmonic intensity is contained near the beam axis, corresponding to the $\ell_{27}$=0 contribution. This is a result of the intensity ratio of the component OAM beams, which has been selected to optimize the channel for on-axis harmonic emission (see Supplemental), such that the surrounding OAM contributions are naturally suppressed (see OAM spectra in the bottom-panels of Fig. 4a). This needle-like beam naturally separates from the driving laser, which has zero on-axis intensity, and can be easily isolated through a pinhole for direct use in an experiment.

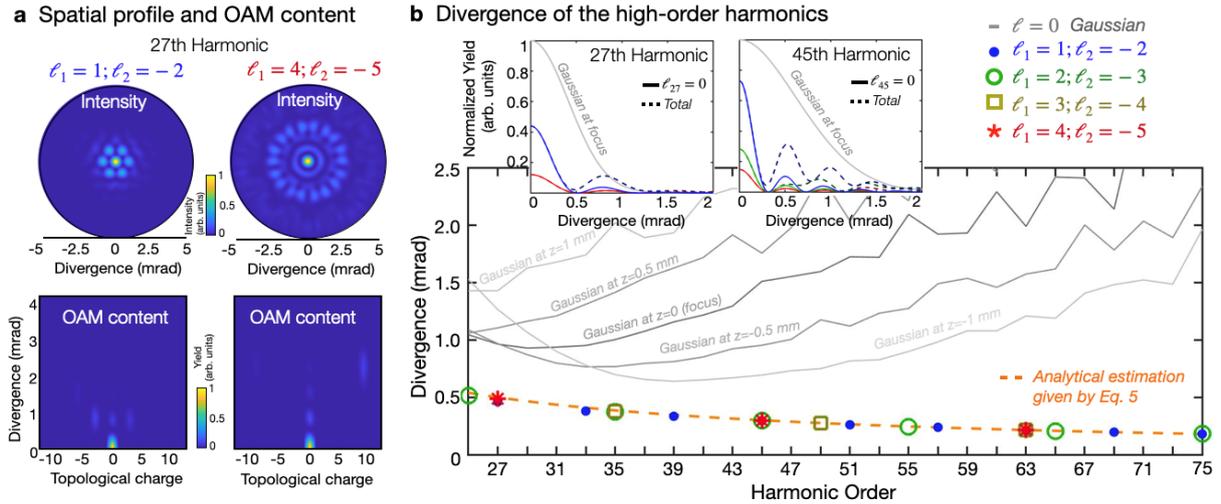

**Figure 4 | Low divergence of the OAM-driven harmonic frequency combs. a** Simulation results of the spatial intensity profile (top) and OAM content (bottom) of the 27$^{th}$ harmonic generated in He for $\ell_1$=1, $\ell_2$=-2 (left) and $\ell_1$=4, $\ell_2$=-5 (right) driving fields. The OAM spectra are obtained from the azimuthal Fourier transform at each divergence angle. **b** Simulation results of the Full width at half maximum (FWHM) divergence of the high-order harmonics for different OAM driving combinations (color dots)—with the gas jet placed at the focus position—, and for a Gaussian driving beam where the gas jet is placed at different positions relative to the focus. The orange dashed line indicates the estimation given by Eq. (5). The top insets show the divergence profile of two sample harmonics (27$^{th}$ and 45$^{th}$) for different OAM driving combinations (showing the $\ell_q$=0 contribution in solid and the total one in dashed line) and for a Gaussian beam where the gas jet is placed at the focus position. Simulation parameters correspond to those of Fig. 2b.

In order to show the low divergence of the $\ell_q$=0 harmonic beam, we present in Fig. 4b the divergence of the high-order harmonics driven by the opposite non-degenerate OAM combination (corresponding to those of Fig. 2b), compared to that of the harmonics driven by a Gaussian beam.

We have chosen a beam waist of 30 μm for the Gaussian beam, as a reasonable comparison against the OAM combination scheme, where the necklace ring presents a radius of 30/√2 μm. The insets show the divergence profile of two sample harmonics (27$^{th}$ and 45$^{th}$) driven by different drivers' OAM combinations (the solid lines show the $\ell$=0 contribution of each harmonic, whereas the dashed lines show the sum of all the OAM contributions), and by a Gaussian beam. Note that the intensity yield of the on-axis emission is comparable to the yield obtained with a Gaussian beam. The bottom panel shows the divergence calculated as the full width at half maximum (FWHM) for different drivers' OAM combinations (dots and dashed lines) —with the gas jet placed at the focus position—and for a Gaussian beam placed at different positions relative to the gas jet (solid grey lines).

It is worth noticing that Gaussian-driven high-order harmonics, counterintuitively, exhibit a divergence that in general increases with the harmonic order (31, 32). In such cases, the divergence of the central intensity peak of the q-th-harmonic scales as $\beta_q \sim \lambda_q/D_q$, $D_q$ being the size of the near-field target area in which the driving field is intense enough to generate that particular harmonic. The fact that $D_q$ decreases with the harmonic order faster than $\lambda_q$ results in an increasing divergence with the harmonic order, as can be observed in Fig. 4b. Remarkably, due to the use of an optimized OAM-driving laser combination, we are able to reverse this behavior. When considering the opposite non-degenerate OAM driving field, the divergence of the $\ell_q$=0 intensity peak of the $q$-th-harmonic scales as

$$\Delta\beta_q^{FWHM} = 2.25 \frac{\lambda_0}{2\pi qR} \quad [5]$$

$R$ being the radius of the necklace driving structure (see Supplemental Material). The result is that higher-order harmonics can be generated with progressively lower divergence, as $R$ is constant for all harmonic orders. Note also that this behavior does not depend on the choice of $\ell_1$ and $\ell_2$, but on the size of the resulting necklace structure, which, for the cases presented in Fig. 4b, presents the same size. As a consequence, similar driving schemes without OAM, such as a Gaussian-necklace driver or a continuous-ring driver, would also result in on-axis high-order harmonics with progressively lower divergence. Note however that our phased-necklace driver is composed on a combination of Laguerre-Gaussian modes, whose propagation behaviour is regular, and which allow us to control the on-axis harmonic content as described in the previous section.

We experimentally validate the predicted divergence behavior of the HHG frequency combs for Ar at 790 nm in Fig. 5. Taking an angular integration centered about the optical axis for the necklace-driven HHG, and considering the imaging condition of the spectrometer, we measure the divergence of the on-axis, Bessel-like lobe. We then compare these divergence values to those obtained with an equivalent Gaussian driver. In order to make as direct a comparison as possible, we match the beam waist parameter (lens focal length), focal position, gas pressure, and peak intensity between the two cases, resulting in HHG spectra with similar spectral envelopes and cutoff photon energies.

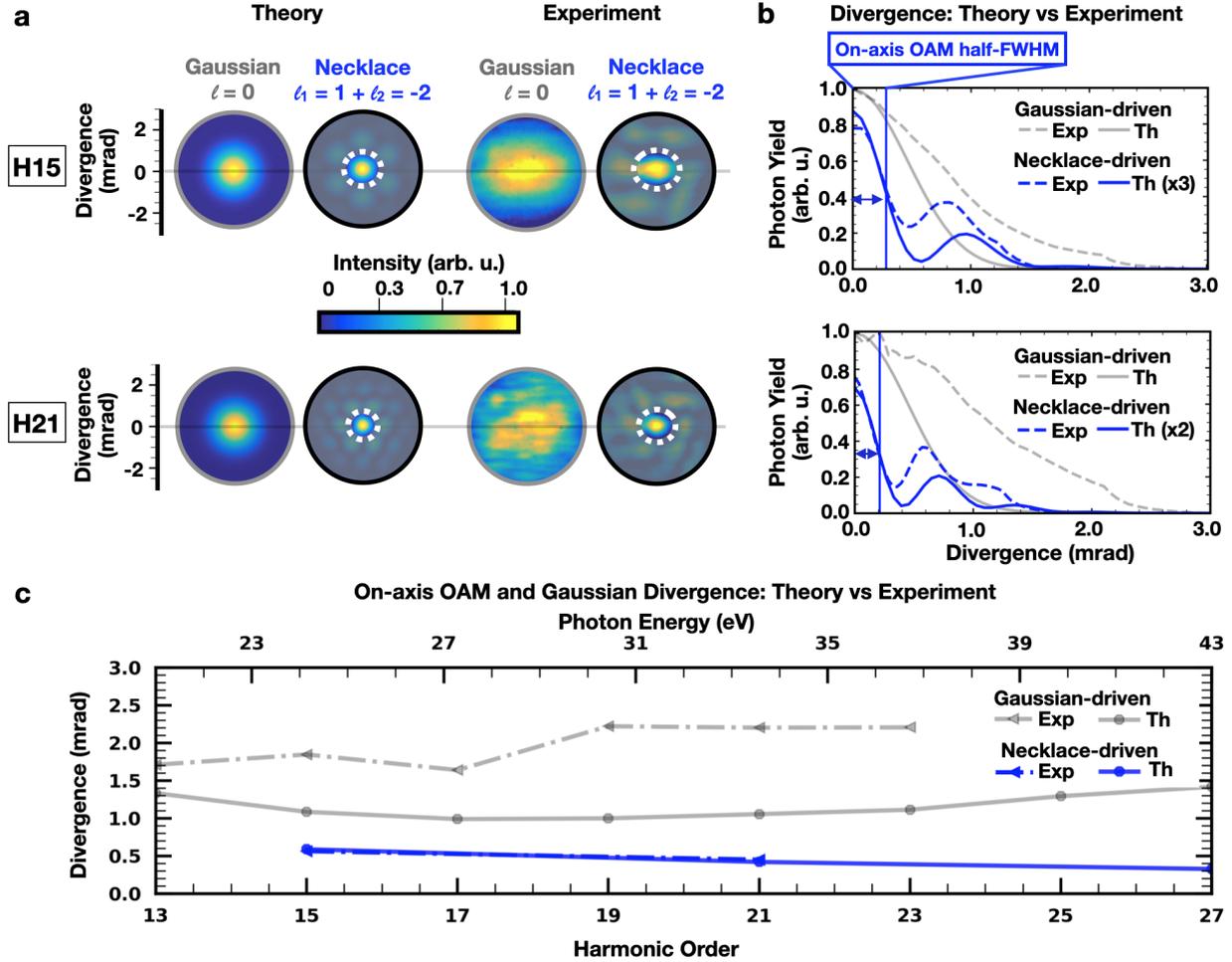

**Figure 5 | Ar-driven harmonic divergences. a** Theoretical (Th.) and experimental (Exp.) comparison of the intensity spatial profiles for the 790-nm Gaussian-driven and necklace-driven ($\ell_1 = 1$ and $\ell_2 = -2$) in Ar gas. The white dashed circles indicate the on-axis emission of the 15$^{\text{th}}$ (H15) and 21$^{\text{st}}$ (H21) for both theory and experiment. We applied an angular integration radially to these profiles to extract the divergences. **b** Angularly integrated radial profiles for H15 and H21 for the on-axis emission, necklace-driven case compared to the equivalent Gaussian. The vertical blue line and double-headed arrow indicate half of the full width at half maximum (FWHM) of the dual-vortex, necklace-driven profile. The intensities of the theoretically predicted on-axis divergence are rescaled for H15 and H21, respectively, to match the intensities of the experimental profiles. **c** Theoretical (Eq. 5) and measured FWHM for necklace-driven on-axis emission profile indicate a decrease in the divergence with increasing harmonic order. This is in contrast to Gaussian-driven HHG, where divergence increases at higher harmonic orders.

From these spectra, we make several observations. First, it is clear that the central lobe of the necklace-driven harmonics with on-axis components has a significantly lower divergence than that achievable with the equivalent Gaussian driver. Furthermore, we observe a clear decrease in the divergence of the 21$^{\text{st}}$ harmonic relative to the 15$^{\text{th}}$ harmonic, whereas the Gaussian harmonics, as expected, actually increase in divergence over the same energy range. Both the divergence values and trend quantitatively agree with our simulations, as well as the prediction given by Eq. (5). Finally, we note that both our numerical and experimental results show that the on-axis harmonics are nearly equal in peak brightness to their Gaussian counterparts, indicating that the frequency selectivity and low divergence achieved in this scheme do not come at the cost of reduced flux at those frequencies. Note that the brightness of the on-axis harmonics can be controlled through the intensity ratio between the drivers (see Supplemental Material).

**Discussion**

Our results present a significant advance for applications that require highly coherent EUV/SXR beams with customizable properties. By changing only the OAM content and radius of the necklace driving field, one can adjust on-demand the line spacing and divergence behavior of the emitted high-harmonic combs in order to tailor the light source for a particular application. Necklace-driven HHG thus enables dramatic control over the HHG spectra without altering the wavelength of the driving laser. The spectral control instead arises as a result of OAM conservation and macroscopic transverse phase-matching conditions—the microscopic HHG physics remains unchanged, thus avoiding tradeoffs in cutoff energy or conversion efficiency. The ability to adjust the spacing between harmonic orders enhances the flexibility of HHG light sources for a variety of applications, particularly in the SXR region, when driven by mid-IR lasers.

In addition to the potential advantages of the unique spectral properties, the needle-like divergence behavior of the harmonics generated in our scheme may enable a simplification of selected imaging and spectroscopy experiments by removing the need for either refocusing optics or methods to eliminate the driving laser beam. Instead, the on-axis contribution of the OAM harmonics could be selected through a pinhole spatial filter, transmitting the full on-axis EUV/SXR flux while blocking the driving laser. This is particularly true at shorter wavelengths due to the unique divergence scaling – since the divergence decreases monotonically with increasing photon energy. This result is especially relevant for HHG driven by mid-infrared lasers, where harmonics up to the $1000^{th}$ can be obtained (see Fig. 2c), for a 2 μm driving laser, extending well into the SXR region. We do not foresee any fundamental restriction to applying this technique to even longer driving wavelengths (22, 50), making it possible to tailor very high-order discrete harmonics with extremely low divergence.

## Conclusion

In conclusion, we have presented a new technique for the generation of EUV and SXR harmonic combs with a line spacing that is controlled through the OAM makeup of infrared necklace driving laser beams. The emitted harmonics exhibit remarkably low divergence which further decreases with the harmonic order, in contrast to standard HHG driven by a Gaussian beam. Our theoretical simulations are corroborated by our experimental results, which demonstrate both the frequency control and divergence behaviour of the generated harmonic combs. Our work will facilitate a variety of applications which require high beam quality and spectroscopic precision, such as tabletop SXR-ARPES or resonant magnetic scattering. The ability to adjust and maintain separation between adjacent harmonic orders is also likely to be beneficial for hyperspectral coherent diffractive imaging techniques (51, 52), particularly in the SXR region. The low divergence and high beam quality enable these experiments to be carried out with simplified setups and enhanced flux throughput. Thus, we believe necklace-driven HHG will become a powerful tool for tabletop EUV/SXR spectroscopy and imaging, as well as measurements of ultrafast charge and spin dynamics at the nanoscale.

## Methods

**Theoretical simulations of HHG driven by a combination of OAM beams.** We use a theoretical method that computes both the full quantum single-atom HHG response and subsequent propagation (49), thus taking into account phase-matching of the high-order harmonics generated in the gas jet. On one hand, the quantum single-atom response is reproduced by calculating the dipole acceleration through the full quantum extended strong field approximation—without performing the saddle-point approximation—, which presents an excellent qualitative and quantitative agreement against the time dependent Schrödinger equation. Such approximation allows us to achieve substantial computational time gain when computing macroscopic HHG over the entire gas jet. On the other hand, harmonic propagation and phase-matching is computed through the electromagnetic field propagator (49). We discretize the target (gas jet) into elementary radiators, assuming that the emitted harmonic radiation propagates with the vacuum phase velocity. In the present simulations, we have assumed an infinitely thin gas jet, flowing along the perpendicular direction to the beam propagation, with a peak pressure of 667 Pa (5 torr). We note that such 2D assumption for the gas target—performed due to computational time limitations— is a reasonable assumption for low density gas jets, based on previous theoretical and experimental results (20, 21). Thus, we do not foresee fundamental deviations if thicker gas jets, closer to the experimental one (150 μm in diameter) are considered.

The spatial structure of the driving beams is represented as a Laguerre-Gaussian beam propagating in the $z$-direction, with wavelength $\lambda_0$, ($k_0 = 2\pi/\lambda_0$), given by

$$LG_{\ell,p}(\rho,\theta,z;k_0) = E_0 \frac{w_0}{w(z)} \left(\frac{\sqrt{2}\rho}{w(z)}\right)^{|\ell|} L_p^{|\ell|}\left(\frac{2\rho^2}{w^2(z)}\right)$$

$$\times \exp\left[-\frac{\rho^2}{w^2(z)}\right] \exp\left[i\ell\theta + i\frac{k_0\rho^2}{2R(z)} + i\phi_G(z)\right] \qquad [2],$$

where $w(z) = w_0\sqrt{1 + (z/z_0)^2}$ is the beam waist ($w_0$ being the beam waist at focus and $z_0 = \pi w_0^2/\lambda_0$ the Rayleigh range), $R(z) = z[1 + (z_0/z)^2]$ is the phase-front radius, $\phi_G(z) = -(2p + |\ell| + 1)\arctan(z/z_0)$ is the Gouy phase, and $L_p^{|\ell|}(x)$ are the associated Laguerre polynomials. $\ell =0,\pm1, ,\pm2,...$ and $p=0,1,2...$ correspond to the topological charge and the number of nonaxial radial nodes of the mode, respectively. In this work we will not consider beams with radial nodes, and thus $p=0$. The driving beam waists ($w_0$) of the different $\ell$ modes are chosen to overlap at the focal plane, which corresponds to $w_0$, $w_0/\sqrt{2}$, $w_0/\sqrt{3}$, $w_0/2$ and $w_0/\sqrt{5}$ for $\ell =\pm1,\pm2, \pm3, \pm4, \pm5$ respectively. In particular, we have considered $w_0 =30$ μm. Finally, the laser pulses are modeled with a trapezoidal envelope. In the simulations performed in Ar and He with 800 nm central wavelength, the envelope consists of three cycles of $\sin^2$ turn-on, ten cycles of constant amplitude—26.7 fs—, and three cycles of $\sin^2$ turn-off. In the simulations performed in He with 2 μm central wavelength, the pulse duration is reduced due to computational time restrictions, and the envelope consists of two cycles of $\sin^2$ turn-on, four cycles of constant amplitude—26.7 fs—, and two cycles of $\sin^2$ turn-off. The amplitudes ($E_0$) of the driving pulses are chosen to obtain a maximum peak intensity at focus at the radii of maximum superposition ($w_0/\sqrt{2}$) of $1.7\times10^{14}$ W/cm$^2$ for Ar, $6.9\times10^{14}$ W/cm$^2$ for He at 800 nm, and $5\times10^{14}$ W/cm$^2$ for He at 2 μm. Note that the intensity ratio for each LG combination has been chosen to optimize the harmonic radiation emitted around the beam axis (see Supplemental Material).

**Experimental generation.** Necklace-beam driven high-harmonics are generated by focusing a pair of collinear, linearly polarized IR-vortex beams (with topological charges of $\ell_1=1$, $\ell_2=-2$) into a supersonic expansion of argon gas. The dual-vortex driver is synthesized from the output of a high-power, ultrafast regenerative amplifier (790 nm, 40 fs, 8 mJ, 1 kHz, KMLabs Wyvern HE) passed through a frequency degenerate Mach-Zehnder interferometer. In each arm of the interferometer, independent spiral phase plates (16 steps per phase ramp, HoloOr), and focusing lenses ($f_1$=40cm, $f_2$=30cm) result in each beam possessing identical (linear) polarization, distinct topological charges, and similarly sized spatial profiles at focus. Independent irises allow for fine tuning of the transverse mode size at the focal plane, and are used to overlap the maximum-intensity ring of the two component beams. The ring of maximum intensity of the two driving beams was matched at a radius of ∼32 μm (see Supplemental Material). Half-waveplate/polarizer pairs are used to independently adjust the pulse energy in each arm in order to optimize the on-axis intensity. For the data presented in this manuscript, the pulse energies are set to $E_1$=480μJ and $E_2$=290μJ. The combined necklace driver is characterized by a modified Gerchberg-Saxton phase retrieval algorithm, which solves for the spatial phase of the composite electric field and allows confirmation of the desired OAM content of the driving IR field, while additionally confirming the high stability of the interferometer setup (see Supplemental). The two component pulses experience approximately equal dispersion throughout each arm of the interferometer, and are confirmed through FROG measurements to have equal pulse widths $\tau \approx$ 57 fs. A high-precision, high-stability translation stage (Newport, XMS-160S) is used to synchronize the two pulses in time. The beams are recombined at the output of the interferometer and focused into the supersonic expansion of argon from a circular gas jet (150μm diameter). The generated high harmonics, which range in photon energy from 20-35 eV, are transmitted through a 200-nm-thick aluminum filter (Luxel), which serves to block the residual IR driver. A removable circular pinhole (200 μm diameter) is placed on the optical axis at a distance of 60cm from the generation region, in order to spatially filter for the on-axis frequency comb. The transmitted harmonics are subsequently focused by a toroidal mirror ($f_{eff}$ = 27cm) and dispersed by a plane ruled EUV grating (1200 grooves/mm, Richardson) at an incidence angle $\boldsymbol{\theta_{inc}} \approx \boldsymbol{45°}$. The high incidence angle is chosen to balance the dispersion and imaging quality of the spectrometer. An EUV CCD camera (Andor Newton 940) is placed behind the toroidal focal plane so as to image the far field (20 cm from the gas jet) of the dispersed harmonics with a magnification of 1.67. These parameters are used to calculate the divergence of the produced harmonics. A 100nm titanium filter (Lebow) with an absorption edge prior to the 21$^{st}$ harmonic (33 eV) is used to verify the photon energies of the harmonics exhibiting on-axis intensity.

## Data Availability

The datasets and analysis routines utilized to prepare the data presented in this manuscript are available, free of charge, from the corresponding author under reasonable request.

# References

[1] J. Miao, T. Ishikawa, I. K. Robinson & M. Murnane. Beyond crystallography: Diffractive imaging using coherent X-ray light sources. *Science* **348**, 530-535 (2015).

[2] Z. Tao, C. Chen, T. Szilvasi, M. Keller, M. Manos, H. Kapteyn, M. Murnane. Direct time-domain observation of attosecond final-state lifetimes in photoemission from solids. *Science* **353**, 62–67 (2016).

[3] J. A. Van Bokhoven & C. Lamberti. *X-Ray Absorption and X-Ray Emission Spectroscopy: Theory and Applications* (John Wiley & Sons: New York, 2016).

[4] Y. Pertot, S. Cédric, M. Matthews, A. Chauvet, M. Huppert, V. Svoboda, A. von Conta, A. Tehlar, D. Baykusheva, J-P Wolf, H. J. Wörner. Time-resolved X-ray absorption spectroscopy with a water window high-harmonic source. *Science* **55**, 264-267 (2017).

[5] Y. Zhang, X. Shi, W. You, Z. Tao, Y. Zhong, F. C. Kabeer, P. Maldonado, P. M. Oppeneer, M. Bauer, K. Rossnagel, H. Kapteyn, M. Murnane. Coherent modulation of the electron temperature and electron-phonon couplings in a 2D material. *PNAS* **117**, 8788-8793 (2020).

[6] P. Emma, R. Akre, J. Arthur, R. Bionta, C. Bostedt, J. Bozek, A. Brachmann, P. Bucksbaum, R. Coffee, F.-J. Decker, Y. Ding, D. Dowell, S. Edstrom, A. Fisher, J. Frisch, S. Gilevich, J. Hastings, G. Hays, Ph. Hering, Z. Huang, R. Iverson, H. Loos, M. Messerschmidt, A. Miahnahri, S. Moeller, H.-D. Nuhn, G. Pile, D. Ratner, J. Rzepiela, D. Schultz, T. Smith, P. Stefan, H. Tompkins, J. Turner, J. Welch, W. White, J. Wu, G. Yocky and J. Galayda. First lasing and operation of an ångstrom-wavelength free-electron laser. *Nat. Photonics* **4**, 641–647 (2010).

[7] T. Ishikawa, H. Aoyagi, T. Asaka, Y. Asano, N. Azumi, T. Bizen, H. Ego, K. Fukami, T. Fukui, Y. Furukawa, S. Goto, H. Hanaki, T. Hara, T. Hasegawa, T. Hatsui, A. Higashiva, T. Hirono, N. Hosoda, M. Ishii, t. Inagaki, Y. Inubushi, T. Itoga, Y. Joti, M. Kago, T. Kameshima, H. Kimura, Y. Kirihara, G. Kiyomochi, T. Kobayashi, C. Kondo, T. Kudo, H. Maesaka, X. M. Maréchal, T. Masuda, S. Matsubara, T. Matsushita, S. Matsui, M. Nagasono, N. Nariyama, H. Ohashi, T. Ohata, T. Ohshima, S. Ono, Y. Otake, C. Saji, T. Sakurai, T. Sato, K. Sawada, T. Seike, K. Shirasawa, T. Sugimoto, S. Suzuki, S. Takahashi, H. Takebe, K. Taheshita, K. Tamasaku, H. Tanaka, R. Tanaka, T. Tanaka, T. Togashi, K. Togawa, A. Tokuhisa, H. Tomizawa, K. Tono, S. Wu, M. Yabashi, M. Yamaga, A. Yamashita, K. Yanagida, C. Zhang, T. Shintake, H. Kitamura & N. Kumagai. A compact X-ray free-electron laser emitting in the sub-ångström region. *Nat. Photonics* **6**, 540–544 (2012).

[8] A. McPherson, G. Gibson, H. Jara, U. Johann, T. S. Luk, I. A. McIntyre, K. Boyer, C. K. Rhodes. Studies of multiphoton production of vacuum-ultraviolet radiation in the rare gases. *J. Opt. Soc. Am. B* **4,** 595 (1997).

[9] M. Ferray, A. L'Huillier, X. F. Li, L. A. Lompre, G. Mainfray, C. Manus. Multiple-harmonic conversion of 1064-nm radiation in the rare gases. *J. Phys. B* **21,** L31 (1998).

[10] A. Rundquist, C. G. Durfee III, Z. Chang, C. Herne, S. Backus, M. M. Murnane, H. C. Kapteyn. Phase-matched generation of coherent soft X-rays. *Science*, **280**, 5368, pp. 1412-1415 (1998).

[11] R. A. Bartels, A. Paul, H. Green, H. C. Kapteyn, M. M. Murnane, S. Backus, I. P. Christov, Y. Liu, A. Attwood, C. Jacobsen. Generation of spatially coherent light at extreme ultraviolet wavelengths. *Science* **297**, 5580, pp. 376-378 (2002).

[12] P. Agostini & L. F. DiMauro. The physics of attosecond light pulses. *Rep. Prog. Phys.* **67** 813–855 (2004).

[13] T. Popmintchev, M.-C. Chen, P. Arpin, M. M. Murnane, H. C. Kapteyn. The attosecond nonlinear optics of bright coherent X-ray generation. *Nat. Photonics* **4** (12), 822–832 (2010).

[14] F. Calegari, G. Sansone, S. Stagira, C. Vozzi & M. Nisoli. Advances in attosecond science. *J. Phys. B* **49**, 062001 (2016).

[15] H. Eichmann, A. Egbert, S. Nolte, C. Momma, B. Wellegehausen, W. Becker, S. Long, J. K. McIver, "Polarization-dependent high-order two-color mixing," *Phys. Rev. A* **51**, R3414–R3417 (1995).

[16] Z. Chang, A. Rundquist, H. Wang, H. Kapteyn, M. M. Murnane. Temporal phase control of soft-X-ray harmonic emission. *Phys. Rev. A Rapid Communications* **58**, R30 (1998).

[17] R. Bartels, S. Backus, E. Zeek, L. Misoguti, G. Vdovin, I. P. Christov, M. M. Murnane, H. C. Kapteyn. Shaped-pulse optimization of coherent emission of high-harmonic soft X-rays. *Nature* **406**, 6792, 164-166 (2000).

[18] T. Fan, P. Grychtol, T. Knut, C. Hernández-García, D. D. Hickstein, D. Zusin, C. Gentry, F. J. Dollar, C. A. Mancuso, C. W. Hogle, O. Kfir, D. Legut, K. Carva, J. L. Ellis, K. M. Dorney, C.

Chen, O. G. Shpyrko, E. E. Fullerton, O. Cohen, P. M. Oppeneer, Dejan B. Milošević, A. Becker, A. Jaroń-Becker, T. Popmintchev, M. M. Murnane, H. C. Kapteyn et al. Bright circularly polarized soft X-ray high harmonics for x-ray magnetic circular dichroism. *PNAS* **112**, 14206–14211 (2015).

[19] D. D. Hickstein, F. J. Dollar, P. Grychtol, J. L. Ellis, R. Knut, C. Hernández-García, D. Zusin, C. Gentry, J. M. Shaw, T. Fan, K. M. Dorney, A. Becker, A. Jaroń-Becker, H. C. Kapteyn, M. M. Murnane, C. G. Durfee. Non-collinear generation of angularly isolated circularly polarized high harmonics. *Nat. Photonics* **9**, 743–750 (2015).

[20] K. M. Dorney, L. Rego, N. J. Brooks, J. San Román, C-T Liao, J. L. Ellis, D. Zusin, C. Gentry, Q. L. Nguyen, J. M. Shaw, A. Picón, L. Plaja, H. C. Kapteyn, M. M. Murnane, C. Hernández-García. Controlling the polarization and vortex charge of attosecond high-harmonic beams via simultaneous spin-orbit momentum conservation. *Nat. Photonics* **13**, 123-130 (2019).

[21] L. Rego, K. M. Dorney, N. J. Brooks, Q. L. Nguyen, C.-T. Liao, J. San Román, D. E. Couch, A. Liu, E. Pisanty, M. Lewenstein, L. Plaja, H. C. Kapteyn, M. M. Murnane, C. Hernández-García. Generation of extreme-ultraviolet beams with time-varying orbital angular momentum. *Science* **364**, eaaw9486 (2019).

[22] T. Popmintchev, M.-C. Chen, D. Popmintchev, P. Arpin, S. Brown, S. Ališauskas, G. Andriukaitis, T. Balčiunas, O. D. Mücke, A. Pugzlys, A. Baltuška, B. Shim, S. E. Schrauth, A. Gaeta, C. Hernández-García, L. Plaja, A. Becker, A. Jaroń-Becker, M. M. Murnane, H. C. Kapteyn. Bright coherent ultrahigh harmonics in the keV X-ray regime from mid-infrared femtosecond lasers. *Science* **336,** 1287 (2012).

[23] D. Popmintchev, C. Hernández-García, F. Dollar, C. Mancuso, J. A. Pérez-Hernández, M.-C. Chen, A. Hankla, X. Gao, B. Shim, A. L. Gaeta, M. Tarazkar, D. A. Romanov, R. J. Levis, J. A. Gaffney, M. Foord, S. B. Libby, A. Jaron-Becker, A. Becker, L. Plaja, M. M. Murnane, H. C. Kapteyn, T. Popmintchev. Ultraviolet surprise: Efficient soft x-ray high-harmonic generation in multiply-ionized plasmas. *Science* **350**, 6265, 1225-1231 (2015).

[24] Z. Chang, A. Rundquist, H. Wang, M. M. Murnane, H. C. Kapteyn. Generation of coherent X-rays at 2.7nm using high harmonic generation. *Phys. Rev. Lett*. **79**, 2967 (1997).

[25] Ch. Spielmann, N. H. Burnett, S. Sartania, R. Koppitsch, M. Schnürer, C. Kan, M. Lenzner, P. Wobrauschek, F. Krausz. Generation of coherent X-rays in the water window using 5-femtosecond laser pulses. *Science* **278**, 661–664 (1997).

[26] E. J. Takahashi, T. Kanai, K. L. Ishikawa, Y. Nabekawa, K. Midorikawa. Coherent water window X ray by phase-matched high-order harmonic generation in neutral media. *Phys. Rev. Lett.* **101**, 253901 (2008).

[27] M.-C. Chen, P. Arpin, T. Popmintchev, M. Gerrity, B. Zhang, M. Seaberg, D. Popmintchev, M. M. Murnane, and H. C. Kapteyn. Bright, coherent, ultrafast soft X-ray harmonics spanning the water window from a tabletop light source. *Phys. Rev. Lett.* **105**, 173901 (2010)

[28] S. M. Teichmann, F. Silva, S. L. Cousin, M. Hemmer, J. Biegert. 0.5-keV Soft X-ray attosecond continua. *Nat. Commun.* **7**, 11493 (2016).

[29] A. S. Johnson, D. R. Austin, D. A. Wood, C. Brahms, A. Gregory, K. B. Holzner, S. Jarosch, E. W. Larsen, S. Parker, C. S. Strüber, P. Ye, John W. G. Tisch, Jon P. Marangos. High-flux soft X-ray harmonic generation from ionization-shaped few-cycle laser pulses. *Sci. Adv.* **4**, eaar3761 (2018).

[30] J. Schötz, B. Förg, W. Schweinberger, I. Liontos, H. A. Masood, A. M. Kamal, C. Jakubeit, N. G. Kling, T. Paasch-Colberg, S. Biswas, M. Högner, I. Pupeza, M. Alharbi, A. M. Azzeer, M. F. Kling. Phase-matching for generation of isolated attosecond XUV and soft-X-ray pulses with few-cycle drivers. *Phys. Rev. X* **10**, 041011 (2020).

[31] H. Wikmark, C. Guo, J. Vogelsang, P. W. Smorenburg, H. Coudert-Alteirac, J. Lahl, J. Peschel, P. Rudawski, H. Dacasa, S. Carlström, S. Maclot, M. B. Gaarde, P. Johnsson, C. L. Arnold, A. L'Huillier. Spatiotemporal coupling of attosecond pulses. *PNAS* **116**, 4779-4787 (2019).

[32] L. Quintard, V. Strelkov, J. Vabek, O. Hort, A. Dubrouil, D. Descamps, F. Burgy, C. Péjot, E. Mével, F. Catoire, E. Constant. Optics-less focusing of XUV high-order harmonics. *Adv. Sci.* **5**, eaau7175 (2019).

[33] H. Rubinsztein-Dunlop, A. Forbes, M. V. Berry, M. R. Dennis, D. L. Andrews, M. Mansuripur, C. Denz, C. Alpmann, P. Banzer, T. Bauer, E. Karimi, L. Marrucci, M. Padgett, M. Ritsch-Marte, N. M. Litchinitser, N. P. Bigelow, C. Rosales-Guzmán, A. Belmonte, J. P. Torres, T. W. Neely, M. Baker, R. Gordon, A. B. Stilgoe, J. Romero, A. G. White, R. Fickler, A. E. Willner, G. Xie, B. McMorran, A. M. Weiner. Roadmap on structured light. *J. Opt.* **19**, 013001 (2017).

[34] R. A. Beth. Mechanical detection and measurement of the angular momentum of light. *Phys.*

*Rev.* **50,** 115-125 (1936).

[35] A. Allen, M. W. Beijersbergen, R. J. C. Spreeuw, J. P. Woerdman. Orbital angular momentum of light the transformation of Laguerre-Gaussian laser modes. *Phys. Rev. A* **45,** 8185-8189 (1992).

[36] M. Zürch, C. Kern, P. Hansinger, A. Dreischuh, Ch. Spielmann. Strong-field physics with singular light beams. *Nat. Phys*. **8**, 743 (2012).

[37] C. Hernández-García, A. Picón, J. San Román, L. Plaja. Attosecond extreme ultraviolet vortices from high-order harmonic generation. *Phys. Rev. Lett.* **111**, 083602 (2013).

[38] G. Gariepy, J. Leach, K. T. Kim, T. J. Hammond, E. Frumker, R. W. Boyd, P. B. Corkum. Creating high-harmonic beams with controlled orbital angular momentum. *Phys. Rev. Lett.* **113**, 153901 (2014).

[39] R. Géneaux, A. Camper, T. Auguste, O. Gobert, J. Caillat, R. Taïeb, T. Ruchon. Synthesis and characterization of attosecond light vortices in the extreme ultraviolet. *Nat. Commun.* **7**, 12583 (2016).

[40] L. Rego, J. San Román, A. Picón, L. Plaja, C. Hernández-García. Nonperturbative twist in the generation of extreme-ultraviolet vortex beams. *Phys. Rev. Lett.* **117,** 163202 (2016).

[41] F. Kong, C. Zhang, F. Bouchard, Z. Li, G. G. Brown, D. H. Ko, T. J. Hammond, L. Arissian, R. W. Boyd, E. Karimi, P. B. Corkum. Controlling the orbital angular momentum of high harmonic vortices. *Nat. Commun*. **8,** 14970 (2017).

[42] D. Gauthier, P. Rebernik Ribic, G. Adhikary, A. Camper, C. Chappuis, R. Cucini, L. F. DiMauro, G. Dovillaire, F. Frassetto, R. Géneaux, P. Miotti, L. Poletto, B. Ressel, C. Spezzani, M. Stupar, T. Ruchon, G. De Ninno. Tunable orbital angular momentum in high-harmonic generation. *Nat. Commun.* **8,** 14971 (2017).

[43] L. Rego, J. San Román, L. Plaja, C. Hernández-García. Trains of attosecond pulses structured with time-ordered polarization states. *Opt. Lett.* **45**, 5636 (2020).

[44] M. Soljačić, S. Sears, M. Segev. Self-trapping of "necklace" beams in self-focusing Kerr media. *Phys. Rev. Lett*. **81**, 4851 (1998).

[45] W. Walasik, S. Z. Silahli, N. M. Litchinitser. Dynamics of necklace beams in nonlinear

colloidal suspensions. *Scientific Reports* **7**, 11709 (2017).

[46] L. Zhu & J. Wang. Arbitrary manipulation of spatial amplitude and phase using phase-only spatial light modulators. *Sci. Rep.* **4**, 7441 (2014).

[47] T. D. Grow, A. A. Ishaaya, L. T. Vuong & A. L. Gaeta. Collapse and stability of necklace beams in Kerr media. *Phys. Rev. Lett*. **99**, 133902 (2007).

[48] J. C. T. Lee, S. J. Alexander, S. D. Kevan, S. Roy & M. J. McMorran. Laguerre-Gauss and Hermite-Gauss soft X-ray states generated using diffractive optics. *Nat. Photon*ics **13**, 205-209 (2019).

[49] C. Hernández-García, C. J. A. Pérez-Hernández, J. Ramos, E. Conejero Jarque, L. Roso, and L. Plaja. High-order harmonic propagation in gases within the discrete dipole approximation. *Phys. Rev. A*. **82**, 0033432 (2010).

[50] C. Hernández-García, J. A. Pérez-Hernández, T. Popmintchev, M. M. Murnane, H. C. Kapteyn, A. Jaron-Becker, A. Becker, and L. Plaja. Zeptosecond high harmonic keV X-ray waveforms driven by midinfrared laser pulses. *Phys. Rev. Lett.* **111**, 2, 033002 (2013).

[51] D. J. Batey, D. Claus & J. M. Rodenburg. Information multiplexing in ptychography. *Ultramicroscopy* **138**, 13–21 (2014).

[52] B. Zhang, D. F. Gardner, M. H. Seaberg, E. R. Shanblatt, C. L. Porter, R. Karl, C. A. Mancuso, H. C. Kapteyn, M. M. Murnane, D. E. Adams. Ptychographic hyperspectral spectromicroscopy with an extreme ultraviolet high harmonic comb. *Opt. Express* **24,** 16 (2016).

[53] M. Abramowitz & I. A. Stegun. *Handbook of Mathematical Functions with Formulas, Graphs and Mathematical Tables* (New York: Dover, 1965).

[54] R. W. Gerchberg & W. O. Saxton. A practical algorithm for the determination of phase from image and diffraction plane pictures. *Optik* **35**, 237-246 (1972).

[55] S. Fu, S. Zhang, T. Wang & C. Gao. Pre-turbulence compensation of orbital angular momentum beams based on a probe and the Gerchberg-Saxton algorithm. *Opt. Lett*. **41**, 3185–3188 (2016).

[56] H. Chang, X.-L. Yin, X.-Z. Cui, Z.-C. Zhang, J.-X. Ma, G.-H. Wu, L.-J. Zhang, X.-J. Xin. Adaptive optics compensation of orbital angular momentum beams with a modified Gerchberg-Saxton-based phase retrieval algorithm. *Opt. Commun*. **405**, 271–275 (2017).

## Acknowledgments

J.S.R., L.P. and C.H.-G. acknowledge support Ministerio de Ciencia e Innovación (FIS2016-75652-P and PID2019-106910GB-I00). This project has received funding from the European Research Council (ERC) under the European Union's Horizon 2020 research and innovation program (Grant Agreement No. 851201). J.S.R., L.P. and C.H.-G. also acknowledge support from Junta de Castilla y León FEDER funds (Project No. SA287P18). L. R. acknowledges support from Ministerio de Educación, Cultura y Deporte (FPU16/02591). C.H.-G. acknowledges Ministerio de Ciencia, Innovación, y Universidades for a Ramón y Cajal contract (RYC-2017-22745), co-funded by the European Social Fund. L.R., J.S.R., L.P. and C.H.-G thankfully acknowledge the computer resources at MareNostrum and the technical support provided by Barcelona Supercomputing Center (FI-2020-3-0013). The JILA team graciously acknowledges support from the Department of Energy BES Award No. DE-FG02-99ER14982 for the experimental implementation, a MURI grant from the Air Force Office of Scientific Research under Award No. FA9550-16-1-0121 for the mid-infrared laser soft X-ray research, and a National Science Foundation Physics Frontier Center grant PHY-1734006 for theory. N.J.B. acknowledges support from National Science Foundation Graduate Research Fellowships (grant no. DGE-1650115). Q.L.D.N. acknowledges support from National Science Foundation Graduate Research Fellowships (grant no. DGE-1144083).

## Author contributions

L.R., J.S.R., L.P., and C.H.-G. performed the theoretical simulations and analyzed the resulting data. N.J.B. and Q.L.D.N. designed and constructed the experiment. N.J.B., Q.L.D.N., and I.B. collected and analyzed experimental data. C.H.-G., L.P., M.M.M., and H.C.K. supervised the theoretical simulations, experimental work, and developed the facilities and measurement capabilities. L.R., N.J.B., Q.L.D.N., J.S.R., M.M.M., L.P., and C.H.-G. wrote and prepared the manuscript. All authors provided constructive improvements and feedback to this work.

## Competing interests

M.M.M. and H.C.K. have a financial interest in KMLabs. The other authors declare no competing financial interests.

# Supplemental Material

# Necklace-structured high harmonic generation for low-divergence, soft X-ray harmonic combs with tunable line spacing

Laura Rego[1]†, Nathan J. Brooks[2]†, Quynh L. D. Nguyen[2]*, Julio San Román[1], Iona Binnie[2], Luis Plaja[1], Henry C. Kapteyn[2], Margaret M. Murnane[2], Carlos Hernández-García[1]

[1]Grupo de Investigación en Aplicaciones del Láser y Fotónica, Departamento de Física Aplicada, University of Salamanca, Salamanca E-37008, Spain

[2]JILA - Department of Physics, University of Colorado and NIST, Boulder, Colorado 80309, USA

†These authors contributed equally to this work.

*Corresponding Authors: Quynh.L.Nguyen@colorado.edu

In this Supplemental material we include information for: (i) Role of the intensity ratio and dipole phase on the brightness of the on-axis emission; (ii) Phase delay between the lobes of the driving necklace; (iii) Theoretical demonstration of the divergence scaling of the on-axis emission; (iv) Experimental characterization of the driving beam; (v) Experimental results for $\ell_1$=2 / $\ell_2$=-3 necklace-driven HHG.

## Role of the amplitude ratio and dipole phase of the drivers to enhance the brightness of the on-axis emission

The relative amplitude or intensity between the two driving fields is a key feature to enhance the brightness of the on-axis high-order harmonic generation (HHG) driven by the opposite, non-degenerate orbital angular momentum (OAM) driving combination. As reported in Ref (40), when driving HHG with two vortex beams, with different topological charges $\ell_1$, $\ell_2$ and overlapped in time and space, each harmonic presents several OAM contributions. The perturbative OAM conservation law implies that the weight of each OAM contribution to the q-th order harmonic is given by the binomial probability distribution associated with the photon number combinations of absorbing $q - n$ photons from mode $\ell_1$ and $n$ photons from mode $\ell_2$. By taking into account that the number of photons absorbed from each mode is proportional to its amplitude, $U_i$, the amplitude ratio between the two driving pulses can be used to shift the center the OAM distribution between $q\ell_1$, and $q\ell_2$. Thus, in order to enhance the on-axis emission ($\ell_q$=0), the amplitude ratio should satisfy $U_1$=($|\ell_2|/|\ell_1|)U_2$. For example, in the case of $\ell_1$=1 and $\ell_2$=-2, for an amplitude ratio of $U_1$=$2U_2$, (intensity ratio of $I_1$=$4I_2$) the OAM distribution of all harmonics will be centered at $\ell_q$=0, and thus the on-axis emission would be the brightest.

However, non-perturbative HHG contains also an OAM signature connected to the intensity dependent harmonic dipole phase, whenever the driving field presents an intensity gradient along the azimuthal coordinate (40). As a result, the harmonic OAM distribution depends not only on the amplitude ratio, but also on the intensity gradient through the dipole phase. therefore, the amplitude ratio that yields the brightest on-axis emission depends on the harmonic order.

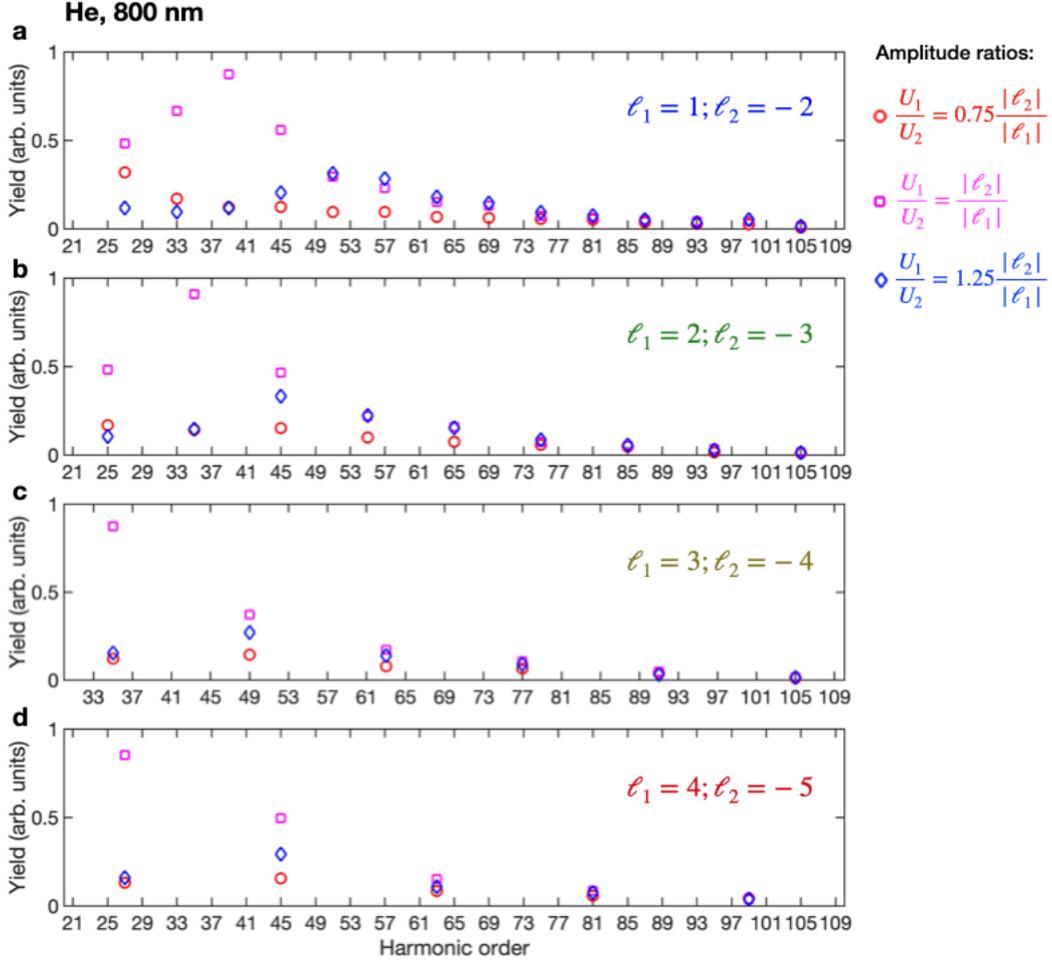

**Figure S1 | Role of the amplitude ratio and dipole phase on the brightness of the on-axis emission.** Simulated on-axis HHG yield driven in He when considering different driving OAM combinations: a) $\ell_1=1$, $\ell_2=-2$, b) $\ell_1=2$, $\ell_2=-3$, c) $\ell_1=3$, $\ell_2=-4$ and d) $\ell_1=4$, $\ell_2=-5$. The amplitudes of the driving pulses are chosen to obtain a maximum peak intensity at focus at the radii of maximum $6.9\times10^{14}$ W/cm$^2$, with intensity ratios of $I_1/I_2=|\ell_2|/|\ell_1|$(red dots), $I_1/I_2=1.5|\ell_2|/|\ell_1|$ (pink dots), and $I_1/I_2=2|\ell_2|/|\ell_1|$ (blue squares). The color bars at the top of each panel indicate the intensity ratio that maximizes the yield of each harmonic. Note that the harmonic yield is normalized for each OAM combination, but it is not normalized between different intensity ratios.

In Figure S1 we present the simulated on-axis HHG yield obtained in He for driving OAM combinations of a) $\ell_1=1$, $\ell_2=-2$, b) $\ell_1=2$, $\ell_2=-3$, c) $\ell_1=3$, $\ell_2=-4$ and d) $\ell_1=4$, $\ell_2=-5$. The amplitudes of the driving pulses are chosen to obtain a maximum peak intensity at focus at the radii of maximum 6.9×10$^{14}$ W/cm$^2$. Three different amplitude ratios have been chosen: $U_1/U_2=0.75|\ell_2|/|\ell_1|$ (red circles), $U_1/U_2=|\ell_2|/|\ell_1|$ (pink squares), and $U_1/U_2=1.25|\ell_2|/|\ell_1|$ (blue diamonds). If the generation process were perturbative, the $U_1/U_2=|\ell_2|/|\ell_1|$ (red dots) case would maximize the on-axis signal for all harmonic orders. However, as it can be observed in all panels from a) to d), the amplitude ratio that optimizes the on-axis harmonic yield varies, especially for the high-order harmonics.

In the numerical simulations presented in the main text, we chose the amplitude (or intensity) ratios $I_1/I_2=1.5|\ell_2|/|\ell_1|$ for all cases presented in Ar at 800 nm. For HHG driven in He at 800 nm, the intensity ratios are $I_1/I_2=4$ for the OAM combination $\ell_1=1$, $\ell_2=-2$; $I_1/I_2=2.25$ for $\ell_1=2$, $\ell_2=-3$; $I_1/I_2=2$ for $\ell_1=3$, $\ell_2=-4$; and $I_1/I_2=1.25$ for $\ell_1=4$, $\ell_2=-5$. Finally, for HHG driven in He at 2 μm, the intensity ratios selected are $I_1/I_2=4$ for the OAM combination $\ell_1=1$, $\ell_2=-2$; $I_1/I_2=3$ for $\ell_1=2$, $\ell_2=-3$; $I_1/I_2=2$ for $\ell_1=3$, $\ell_2=-4$; and $I_1/I_2=1.88$ for $\ell_1=4$, $\ell_2=-5$.

## **Temporal cadence and relative phase of the harmonic emission from the necklace lobes.**

The necklace-driven high-order harmonics detected on-axis are emitted as a sequence of $\Delta q = \Delta\omega/\omega_0$ bursts per cycle of the driving pulse (see Eq. (3) in the main text). This is a result expected from the Fourier transform of the generated harmonic comb, which is a direct consequence of the coherent macroscopic emission of the necklace-driven harmonics. The different lobes in the necklace driving field possess different phases, which implies that the harmonics are emitted at different instants in each lobe. Since the on-axis emission is detected at the same distance from all lobes, the minimum phase shift between the lobes determines the time delay between consecutive on-axis HHG bursts.

Let us consider the necklace driving field composed by two OAM components $\ell_1 = |\ell_1|$ and $\ell_2 = -|\ell_2|$, with $|\ell_1| < |\ell_2|$. To understand the azimuthal field structure, let us first consider that the two OAM beams overlap at their radius of maximum intensity so the field at that radius can be written as,

$$U(\theta) = U_1 e^{i|\ell_1|\theta} + U_2 e^{-i|\ell_2|\theta}, \qquad [S1]$$

where $\theta$ is the azimuthal coordinate and $U_i$ representes the amplitude of each OAM component. From $|U(\theta)|^2$ we find that there are $N = |\ell_1| + |\ell_2|$ lobes in the necklace, located at $\theta_n = n\,2\pi/N$, $n$ ranging from 0 to $N$-1. By writing the field at the radius of maximum intensity as $U(\theta) = U_1 e^{i|\ell_1|\theta}(1 + (U_2/U_1)e^{-iN\theta})$, we can calculate the phase of the driving field at the position of maximum intensity of each lobe:

$$\phi(\theta_n) = n\,2\pi \frac{|\ell_1|}{|\ell_1|+|\ell_2|}, \qquad n = 0,1,\ldots,N-1 \qquad [S2]$$

Note that this phase is equivalent to $\phi(\theta_n) = n\,2\pi \frac{-|\ell_2|}{|\ell_1|+|\ell_2|}$.

By inspecting the above expression one can find particular OAM combinations in which there are lobes that exhibit the same phase, and, thus, their emission is constructively superimposed in time in the on-axis signal. Therefore, to find the number of harmonic bursts emitted, it is necessary to determine the number of distinguishable lobes of the driving field. It must be a fraction of the number of lobes, $N_d = \frac{(|\ell_2|+|\ell_1|)}{K}$, $K$ being a natural number that indicates the number of repetitions of each phase value or, in other words, the number of lobes that are indistinguishable for each OAM combination. $K$ can be calculated as the number of azimuthal angles in which the $\ell_1$-component is in phase with the $\ell_2$-component, which is the greatest common divisor (g.c.d.) between $|\ell_1|$ and $|\ell_2|$: $K = g.c.d(|\ell_1|,|\ell_2|)$. From the mathematical relation $g.c.d(a,b) = \frac{a \cdot b}{l.c.m(a,b)}$, being l.c.m. the least common multiple, we can express $K = \frac{|\ell_1||\ell_2|}{l.c.m(|\ell_1|,|\ell_2|)}$ for convenience. Following the same notation as in the main text—where we introduced $\xi_1 = L_{lcm}/|\ell_1|$ and $\xi_2 = L_{lcm}/|\ell_2|$, $L_{lcm}$ being the least common multiple of $|\ell_1|$ and $|\ell_2|$—, the number of distinguishable lobes is $N_d = \frac{(|\ell_2|+|\ell_1|)}{|\ell_2||\ell_1|} L_{lcm} = \xi_1 + \xi_2$.

Let us now calculate the minimum phase shift between two of these distinguishable lobes, $\Delta\phi_{min}$. First, it is necessary to recall, that the HHG emission occurs twice per cycle as a consequence of the symmetry of the driving field and atomic target, which means that the fundamental phase shift between the emission of two harmonic bursts in each lobe is $\omega_0 T_0/2 = \pi$, where $\omega_0$ is the driving field frequency and $T_0$ its corresponding oscillation period. As a consequence of this periodicity, we are interested in the modulo $\pi$ of the lobe phases. Second, note that the variation of the driving field phase at the peak intensity of the lobes is constant along the azimuthal angle (see Eq. [S2]). Therefore, since the number of distinguishable lobes is $N_d$ and their phases must range linearly from 0 to $\pi$, the minimum phase shift between different lobes is $\Delta\phi_{min} = \frac{\pi}{N_d} = \frac{\pi}{\xi_1+\xi_2}$.

Finally, from the minimum phase shift—which may occur between non-consecutives lobes—, we can extract the time delay between successive emissions, $\Delta\tau$:

$$\Delta\phi_{min} = \omega_0 \Delta\tau \rightarrow \Delta\tau = \frac{\Delta\phi_{min}}{\omega_0} = \frac{T}{2(\xi_1+\xi_2)}. \quad [S3]$$

Therefore, the number of bursts per cycle are $2(\xi_1 + \xi_2)$, which is equal to the line spacing of the harmonic comb $\Delta\omega/\omega_0$ (Eq. (3) in the main text).

## **Divergence of the on-axis HHG emission driven by two opposite non-degenerate OAM driving beams**

To estimate the divergence of the harmonic beams emitted on axis we calculate their far-field distribution. Assuming that the qth-harmonic presents on-axis signal, we consider that its spatial distribution at the gas target (the harmonic near-field distribution) can be approximated as a thin ring (which would correspond to the ring of maximum intensity) with an azimuthal dependence related to the OAM content of that harmonic:

$$A_q(\rho', \phi', z=0) = A_0 \delta(\rho' - R) \sum_{\ell=-\infty}^{\infty} c_\ell e^{i\ell\phi'}, \quad [S4]$$

where $R$ is the radius of the ring and $c_\ell \neq 0$ only for the OAM values that contribute to the harmonic emission. If we introduce this near-field distribution in the Fraunhofer integral formula, we end up with the following expression:

$$U_q(\rho, \phi, z) \propto A_0 R \int_0^{2\pi} \left( \sum_{\ell=-\infty}^{\infty} c_\ell e^{i\ell\phi'} \right) e^{-i\frac{2\pi}{\lambda_q z}\rho R \cos(\phi-\phi')} d\phi' \quad [S5]$$

Making use of the Jacobi-Anger identity (53), we expand the exponential of the trigonometric function in basis of their harmonics obtaining two different contributions to the far-field spatial distribution: $U_q(\rho, \phi, z) = U_q^{(0)}(\rho, \phi, z) + U_q^{(1)}(\rho, \phi, z)$, where:

$$U_q^{(0)}(\rho, \phi, z) \propto A_0 R J_0\left(\frac{2\pi}{\lambda_q z}\rho R\right) \int_0^{2\pi} \left( \sum_{\ell=-\infty}^{\infty} c_\ell e^{i\ell\phi'} \right) d\phi' = 2\pi R J_0\left(\frac{2\pi}{\lambda_q z}\rho R\right) c_{\ell=0} \quad [S6]$$

$$U_q^{(1)}(\rho, \phi, z) \propto A_0 2R \sum_{n=1}^{\infty} (-i)^n J_n\left(\frac{2\pi}{\lambda_q z}\rho R\right) \int_0^{2\pi} \left( \sum_{\ell=-\infty}^{\infty} c_\ell e^{i\ell\phi'} \right) \cos(n(\phi-\phi')) d\phi' \quad [S7]$$

The first contribution, $U_q^{(0)}(\rho, \phi, z)$, is the only one that exhibits signal on-axis and, as shown in Eq. [S6], and it is related to the zero OAM contribution of the q-th order harmonic, $c_{\ell=0}$. This result is expected, as the q-th order harmonic exhibits on-axis signal only if it possesses $\ell=0$ content. We can now calculate its divergence distribution as:

$$U_q^{(0)}(\beta) \sim J_0\left(\frac{2\pi R}{\lambda_q}\beta\right), \quad [S8]$$

where $\beta \sim \rho/z$. Moreover, we can calculate the divergence width of the qth-harmonic. The Full Width at Half Maximum (FWHM) divergence angle of the intensity distribution is:

$$\Delta\beta_q^{FWHM} = 2.25 \frac{\lambda_q}{2\pi R} = 2.25 \frac{\lambda_0}{2\pi q R}. \quad [S9]$$

This estimation agrees very well with the results obtained in the experiments and simulations presented in the main text, demonstrating that the divergence angle of the harmonics emitted on-axis decreases with the harmonic order. Moreover, the estimation shows that the larger the harmonic ring is, the sharper the comb of harmonics emitted on-axis becomes. To test the dependence of the divergence of the harmonics emitted on axis vs. the size of the ring, we have performed simulations changing the radius of the necklace ring of the driving beam and, therefore, the size of the ring of the EUV/SXR phased antenna array. Figure S2 shows the FWHM divergence of the harmonics emitted on axis for the case of a driving beam composed of $\ell_1$=2 and $\ell_2$=-3 vortex beams, considering different ring radii, *R*, (28.3 μm, 21.2 μm and 14.1 μm). We have also added in the figure the estimation indicated by Eq. [S9] showing a very good agreement with those obtained in the full-quantum simulations, which demonstrates the decrease of the divergence when increasing the size of the vortices used as driving beam.

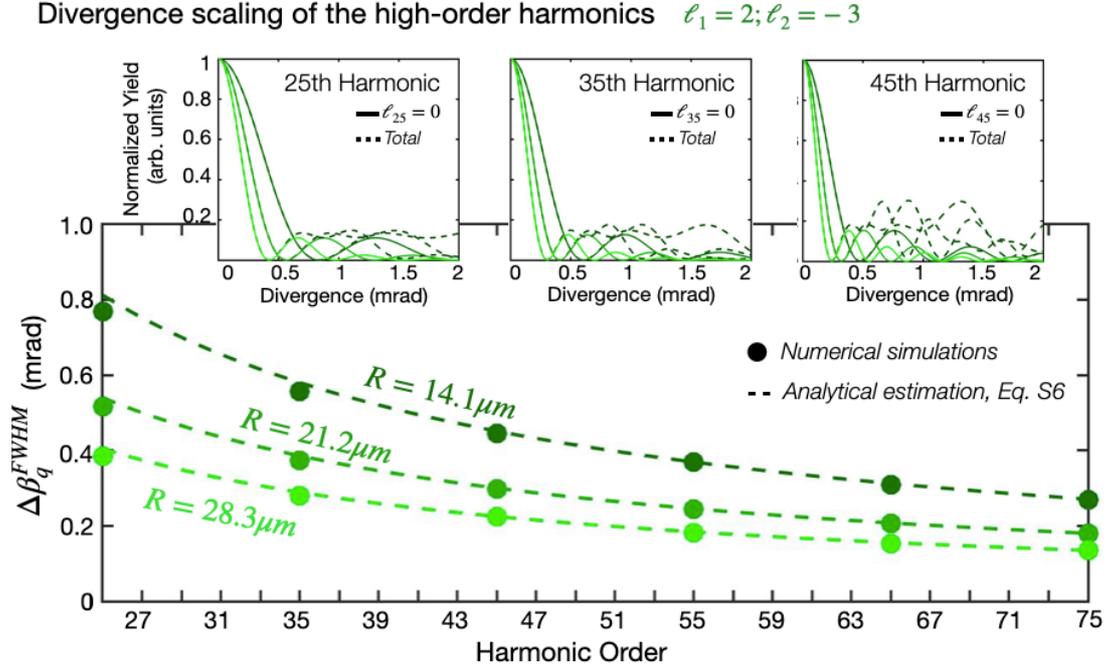

**Figure S2 | Divergence of the OAM-driven high harmonic combs for different driving beam waists.** Full width at half maximum (FWHM) divergence of the high-order harmonics for a driving beam composed of $\ell_1$=2 and $\ell_2$=-3 vortex beams, with beam waists adjusted so the radius of the ring of maximum intensity (R) is 28.3 μm (dark green), 21.2 μm (regular green) and 14.1 μm (light green). Those values of R correspond to $w_0$ of 20, 30 and 40 μm respectively (see Methods). The dots correspond to the numerical simulations whereas the dashed lines to the analytical estimation given by Eq. [S9]. The top insets show the divergence profiles of three sample harmonics (25th, 35th and 45th), showing the $\ell_q$=0 contribution in solid and the total one in dashed line.

## **Experimental characterization of the necklace driving beam composed of two opposite non-degenerate OAM fields**

The necklace beams described in this work can be generated by the interference of two vortex beams each carrying pure, opposite and non-degenerate orbital angular momenta. Experimentally, we form a three-lobe necklace driver for HHG by interferometrically combining two ultrafast pulses with OAM $\ell_1 = 1$ and $\ell_2 = -2$. In order to stably produce the predicted low divergence, on-axis high harmonic comb, we take care to ensure high spatiotemporal quality of the component pulses, as well as excellent relative temporal and spatial stability.

Using a partial reflection from a wedge inserted at the output of the interferometer, we observe the combined driver using a visible CCD (Mightex BTE-B050-U) mounted on a translation stage.

The intensity of the combined field can thus be observed at various $z$-positions, shown in Fig. S3a ($z$ = 0mm) and Fig. S3b ($z$ = 9mm), showing that the necklace structure persists in and out of the focal plane. The component pulses are synchronized in time by maximizing the interference contrast in these images, and confirmed to have identical pulse durations ($\tau_{FWHM}$ = 57 fs) by individual FROG measurements (Fig. S3c) – here, the relative amplitudes are plotted to reflect the intensity ratio of the component pulses, which is set to optimize the on-axis HHG emission.

A series of seven images are acquired at various $z$-positions on either side of and including the focal plane. These images form an overconstrained data set corresponding to a unique solution for the complex electric field (phase and amplitude) of the beam. This solution is obtained by feeding the images into a Gerchberg-Saxton phase-retrieval algorithm (54-56). Beginning with a random guess for the spatial phase profile, the algorithm uses numerical Fresnel propagation in order to enforce consistency with the measured images at each $z$-position. By iteratively cycling through the acquired image stack, the algorithm converges to the correct phase within 20 iterations. For the present combination ($\ell = 1$, $\ell = -2$), the reconstruction shows that each of the lobes in the necklace has approximately constant phase in the focal plane, with the expected relative phase shift between neighbors of $2\pi/3$ (Fig. S3d).

We integrate the amplitude profile of the beam at focus about the azimuth in order to measure the radius of maximum intensity $R \approx 32$ μm (Fig. S3e), which can be used to calculate the expected divergence of the on-axis HHG comb. We then compare this to the azimuthal Fourier transform of the retrieved complex electric field, which gives the OAM distribution as a function of radius. We see that at the necklace radius $R$, the two OAM channels have an amplitude ratio of $|E_1|/|E_2| \approx 1.72$, consistent with the theory prediction for optimized on-axis emission.

We note that in addition to confirming the expected phase structure and OAM content of the necklace beam, the success of the phase reconstruction also indicates the high degree of stability of our interferometer. The phase-retrieval algorithm used here is suited for measuring the phase of a quasi-monochromatic, spatially and temporally coherent beam. In order to apply it for the superposition of multiple pulses, a high degree of mutual coherence is necessary. Beyond the necessary conditions of identical wavelength and polarization, significant spatial or temporal jitter between the component pulses causes the acquired images to morph and/or smear. The images would then be inconsistent with a naturally diffracting field with the expected properties, and cause the reconstruction to fail. Hence, by applying the GS technique to the total (combined) electric field, we confirm the quality of both the necklace beam and the interferometer setup used to form it.

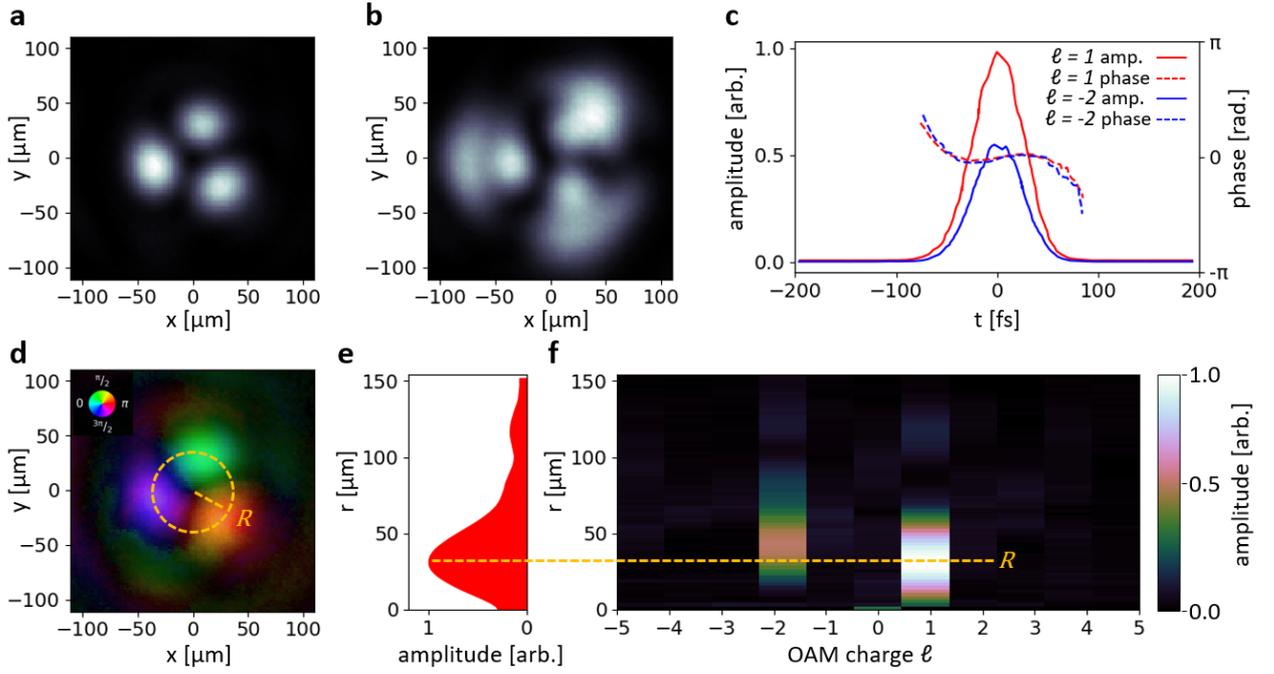

**Figure S3 | Experimental Characterization of Dual-Vortex Necklace Driving Beam.** Images are acquired of the combined 800nm ($\ell = 1$, $\ell = -2$) driver at several points in (a) and out of (b) the laser focus using a visible CCD. The amplitude of the $\ell = 1$ pulse is increased relative to that of the $\ell = -2$ pulse in order to optimize the on-axis emission in the generated harmonics. The pulses are set to be temporally overlapped by maximizing the interference contrast in the CCD images, and confirmed to have identical pulse durations and structure via individual FROG measurements (c). The CCD images of the driver are used in a Gerchberg-Saxton phase-retrieval algorithm to measure the complex electric field of the combined beam (d), where brightness and color correspond to amplitude and phase, respectively. Angular integration of the necklace-structured electric field amplitude (e) gives the radius of maximum intensity $R \approx 32$ μm, while an azimuthal Fourier transform of the complex electric field (f) gives the OAM content, including the relative strength and approximately matched radial positions.

## Experimental results for $\ell_1$=2 / $\ell_2$=-3 necklace-driven harmonics

We experimentally corroborate the selection rules describing necklace-driven HHG for the case of a driving beam composed of OAM $\ell_1$=2 and $\ell_2$=-3. In order to generate the $\ell_2$=-3 beam with our available set of phase optics, we modified one of the arms of our interferometer (Fig. S4a). After reflecting from a thin film polarizer (TFP) pair, the beam passes through a series of optics including a quarter wave plate (QWP) set at 45°, a 790 nm/$\ell$ = 1 spiral phase plate (SPP), and a 390 nm/$\ell$ = 1 SPP, approximately equivalent to a 790nm/$\ell$ = 0.5. After normal incidence reflection from a dielectric mirror (switching the sign of the OAM charge), the beam passes through this set of optics in reverse, thus accumulating a total of $\ell$ = 2 x -(1 + 0.5) = -3 from the SPPs, and a 90° polarization rotation from the QWP. The resultant $\ell$ = -3 beam then transmits through the TFP, after which it is restored to its original polarization with a half wave plate (HWP). Finally, it is combined with an ℓ = 2 arm to form a 5-lobed necklace beam with the expected transverse phase structure in the focal plane (Fig. S4b).

As with the $\ell_1=1$, $\ell_2=-2$ case, we use this necklace beam to drive HHG in argon. Within the bandwidth of harmonics produced, only the 15th harmonic order (H15) is predicted by OAM selection rules/transverse phase matching conditions to have on-axis intensity. This is supported by looking at the full harmonic spatial profiles, as H15 shows a distinct maximum in the center which is not present for other "forbidden" harmonics orders such as H19 (Fig. S4c). Inserting a pinhole onto the optical axis transmits H15 while relatively suppressing the other harmonic orders (Fig. S4c inset, S4d), confirming the on-axis selection rules for this case. The residual intensity at the forbidden harmonics is likely the result of an imperfect ℓ = -3 mode resulting from the complex generation method described above, as well as slight misalignment of the pinhole relative to the optical axis.

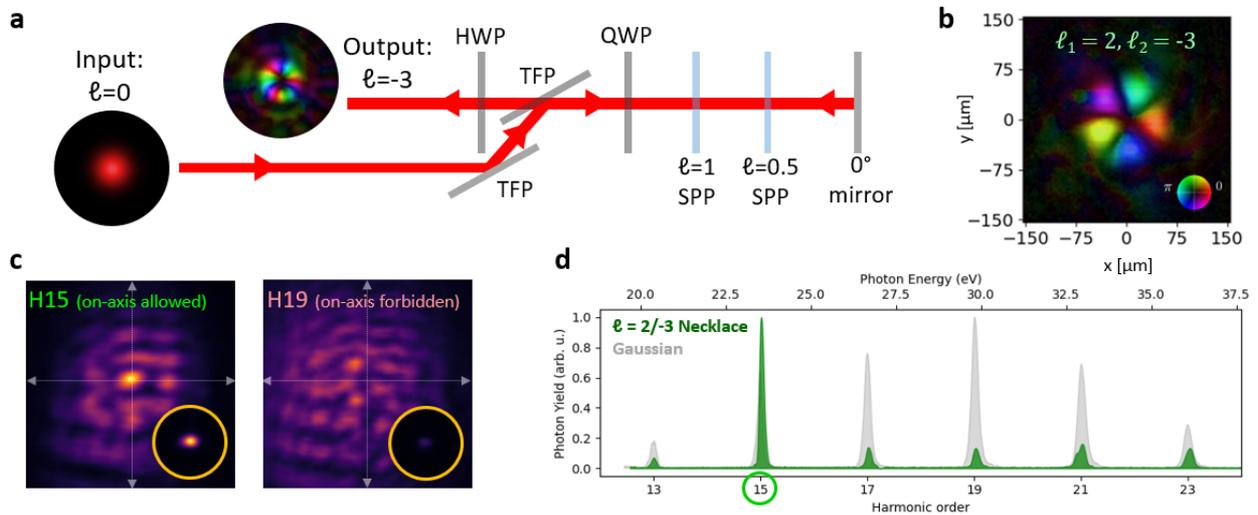

**Figure S4 | Experimental results for $\ell_1$=2 and $\ell_2$=-3 necklace-driven harmonics.** a) Experimental setup for creating ℓ=-3 beam. The output beam is combined downstream with the ℓ=2 arm to create the five lobed necklace beam in b), shown with experimentally retrieved phase profile in the focal plane, and used to drive harmonics in argon. c) Full spatial profiles of emitted 15th (on-axis allowed) and 19th (on-axis forbidden) harmonic orders, with transmission through an on-axis pinhole shown in the inset. d) Full spectrum transmitted through on-axis pinhole compared to Gaussian HHG.

9